\documentclass[aps,prc,10pt,amsmath,amssymb,twocolumn,floatfix,nofootinbib,superscriptaddress,hyperref]{revtex4-1}

\usepackage{hyperref}
\usepackage{graphicx,color,rotating,pifont}
\usepackage{epstopdf}
\usepackage{amsmath,amssymb,bm}
\usepackage{ae}
\usepackage{pstricks}
\usepackage{dcolumn}
\usepackage{tikz}
\usepackage{bbold}
\DeclareMathSymbol{\NS}{\mathord}{AMSb}{"4E}

\newcommand{\totd}[2]{\ensuremath{ \frac{d {#1}} {d {#2}} }}

\definecolor{FGViolet}{rgb}{0.61,0.32,0.61}
\definecolor{FGDarkBlue}{rgb}{0,0,0.6}
\definecolor{FGBlue}{rgb}{0,0,0.8}
\definecolor{FGLightBlue}{rgb}{0.2, 0.6, 0.8}
\definecolor{FGGreen}{rgb}{0.2,0.7,0.2}
\definecolor{FXGreen}{rgb}{0.2,0.7,0.4}
\definecolor{FGLightGreen}{rgb}{0.4,1,0.4}
\definecolor{FGYellow}{rgb}{1,0.95,0}
\definecolor{FGOrange}{rgb}{0.95,0.5,0.1}
\definecolor{FGRed}{rgb}{0.8,0,0}
\definecolor{FGWhite}{rgb}{1,1,1}
\definecolor{FGLightGray}{rgb}{0.8,0.8,0.8}
\definecolor{FGGray}{rgb}{0.5,0.5,0.5}
\definecolor{FGDarkGray}{rgb}{0.3,0.3,0.3}
\begin{document}
\title{Ab Initio Excited States from the In-Medium Similarity Renormalization Group}
\author{N. M.\ Parzuchowski}
\email[E-mail:~]{parzuchowski@frib.msu.edu}
\affiliation{National Superconducting Cyclotron Laboratory
and Department of Physics and Astronomy, Michigan State University,
East Lansing, Michigan 48824, USA}
\author{T. D.\ Morris}
\email[E-mail:~]{tmorri31@utk.edu}
\affiliation{Department of Physics and Astronomy, University of Tennessee,
Knoxville, TN 37996, USA} 
\affiliation{Physics Division, Oak Ridge National Laboratory,
Oak Ridge, TN 37831, USA} 
\affiliation{National Superconducting Cyclotron Laboratory
and Department of Physics and Astronomy, Michigan State University,
East Lansing, Michigan 48824, USA}
\author{S. K.\ Bogner}
\email[E-mail:~]{bogner@nscl.msu.edu}
\affiliation{National Superconducting Cyclotron Laboratory
and Department of Physics and Astronomy, Michigan State University,
East Lansing, Michigan 48824, USA}
\date{\today}
\begin{abstract}

We present two new methods for performing \emph{ab initio} calculations of excited states for closed-shell systems within the in-medium similarity renormalization group (IMSRG) framework. Both are based on combining the IMSRG with simple many-body methods commonly used to target excited states, such as the Tamm-Dancoff approximation (TDA) and equations-of-motion (EOM) techniques. In the first approach, a two-step sequential IMSRG transformation is used to drive the Hamiltonian to a form where a simple TDA calculation (i.e., diagonalization in the space of $1$p$1$h excitations) becomes exact for a subset of eigenvalues.  In the second approach, equations-of-motion (EOM) techniques are applied to the IMSRG ground-state-decoupled Hamiltonian to access excited states. We perform proof-of-principle calculations for parabolic quantum dots in two-dimensions and the closed shell nuclei $^{16}$O and $^{22}$O.  We find that the TDA-IMSRG approach gives better accuracy than the EOM-IMSRG when calculations converge, but is otherwise lacking the versatility and numerical stability of the latter. Our calculated spectra are in reasonable agreement with analogous EOM-coupled-cluster (EOM-CC) calculations. This work paves the way for more interesting applications of the EOM-IMSRG to calculations of consistently evolved observables such as electromagnetic strength functions and nuclear matrix elements, and extensions to nuclei within 1-2 nucleons of a closed shell by generalizing the EOM ladder operator to include particle-number nonconserving terms. 
\end{abstract}
\pacs{??}
\maketitle
\clearpage
\section{Introduction\label{sec:intro}}

As experimental efforts have shifted towards the study of rare isotopes, there has been an increased demand for reliable \emph{ab initio} calculations to counter the inherent limitations of phenomenological approaches.  For decades \emph{ab initio} progress in  theory was slowed by the lack of a consistent theory for the strong inter-nucleon interactions, and by the computational demands required to handle the non-perturbative aspects of the problem.  For many years, the only option for controlled calculations was to use quasi-exact
methods such as quantum Monte Carlo (QMC)~\cite{Pudliner:1997ck,Pieper:2001mp,Carlson:2015lq} or no-core shell model (NCSM)~\cite{Navratil:2000gs,Navratil:2009ut,Barrett:2013oq} methods, which limited the reach of \emph{ab initio} calculations to light $p$-shell nuclei. Approximate (but systematically improvable) methods that scale favorably to larger systems, like coupled cluster (CC) theory and many-body perturbation theory (MBPT), were largely abandoned in nuclear physics, despite enjoying tremendous success in quantum chemistry~\cite{Shavitt:2009}.

Impressive progress has been made in recent years as advances in chiral effective field theory (EFT)~\cite{Epelbaum:2009ve,Machleidt:2011bh} which provides a systematic framework to construct consistent two- and three-nucleon interactions, and the increasing prevalence of powerful renormalization group (RG) methods~\cite{Bogner:2010pq,Roth:2011kx}, which enable one to transform interactions to much softer forms, have led to a resurgence of CC and similar methods such as self-consistent Green's functions (SCGF) and the in-medium similarity renormalization group (IMSRG), pushing the frontiers of \emph{ab initio} theory well into the medium-mass region~\cite{Tsukiyama:2011uq,Tsukiyama:2012fk,Bogner:2014tg,Jansen:2014qf,Jansen:2015ngw,Stroberg:2015ymf,Stroberg:2016ung,Soma:2013ys,Soma:2014fu,Soma:2014eu,Hergert:2013ij,Hergert:2014vn,Hergert:2016etg, Wienholtz:2013bh,Hagen:2015ve}. Early applications of these methods were limited primarily to ground-state properties of stable nuclei near shell closures with two-nucleon forces only. Substantial progress has since been made on including three-nucleon forces~\cite{Hagen:2007zc,Soma:2013xha,Roth:2012qf, Hergert:2012nb}, targeting excited states and observables besides energy~\cite{Ekstrom:2014iya,Jansen:2012ey}, and moving into the more challenging terrain of open-shell and unstable nuclei~\cite{Tsukiyama:2012fk,Jansen:2014qf,Bogner:2014tg,Stroberg:2015ymf,Stroberg:2016ung,Soma:2013ys,Soma:2012zd,Hergert:2013ij,Gebrerufael:2016xih}. Remarkably, progress on the many-body front has been so swift in recent years that inadequacies of the current-generation chiral two- and three-nucleon interactions, rather than the many-body calculations themselves, are the primary obstacles to systematic calculations across the medium-mass region~\cite{Ekstrom:2015fk,Binder:2014fk}.

The IMSRG framework is particularly appealing because it offers several paths to calculate ground and excited state properties for closed- and open-shell systems. One promising approach for open-shell nuclei is to use the IMSRG to construct a valence-space Hamiltonian that is decoupled from the much larger Hilbert space of the full $A$-body problem, which is then diagonalized using standard shell model machinery. Initial applications in the $sd$-shell have been quite encouraging, giving a marked improvement over previous valence-space Hamiltonians constructed in MBPT, and clearly demonstrating the importance of three-nucleon interactions in reproducing experimental spectra~\cite{Bogner:2014tg,Stroberg:2015ymf,Stroberg:2016ung}.  

The valence-space decoupling has the virtue of providing a unified treatment of ground- and excited-state properties (including deformation and transitions) couched in the familiar language of the phenomenological shell model, but it also suffers the same ``curse of dimensionality'' associated with the large-scale matrix diagonalizations that are required to access midshell nuclei and/or extended valence spaces.  One alternative that bypasses these difficulties is to directly target excited states by combining the IMSRG with equations-of-motion (EOM) techniques~\cite{Rowe:1968eq}, similar to what is done in CC theory~\cite{Ekstrom:2014iya,Jansen:2012ey}. While the EOM-IMSRG potentially offers some technical simplifications due to the Hermiticity of the transformed Hamiltonian (e.g., no need to solve a separate left-eigenvalue problem when calculating properties other than energy), the practical limitations of the single-reference formulation should be comparable to the analogous EOM-CC calculations, limiting the method to nuclei within 1 or 2 nucleons of a closed shell.  

To remove these limitations, one possibility is to merge EOM techniques with the multi-reference IMSRG (MR-IMSRG) formulation recently developed for ground-state calculations of open-shell even-even nuclei~\cite{Hergert:2013ij,Hergert:2014vn}. In principle, spectroscopy for the target nucleus and its even-odd, odd-even, and odd-odd neighbors could then be accessed using suitably generalized EOM excitation operators. Since the full implementation of the MR-EOM-IMSRG is a significant undertaking, we first develop the single-reference EOM-IMSRG to calculate excited states in closed-shell systems as a ``proof-of-principle'', before taking on the much more challenging multi-reference formulation.  In the following we will show that the EOM-IMSRG is indeed a viable approach to target excited states, giving good agreement with analogous EOM-CC calculations for the $^{16}$O and $^{22}$O nuclei considered, and exhibiting systematic improvement towards the exact full configuration interaction (FCI) results in 2d quantum dots when perturbative triple-excitation corrections are included in our EOM calculations. 

This work is organized as follows. In Section~\ref{sec:formalism}, we review the basic formalism of the IMSRG and present two different strategies for targeting excited states based on i) sequentially transforming the Hamiltonian to a block-diagonal form in particle-hole excitations and then diagonalizing the 1-particle 1-hole block of the transformed Hamiltonian, and ii) performing an EOM calculation with single- and double-excitation operators using the ground-state-decoupled Hamiltonian.  For the latter, we also present a simple perturbative procedure that corrects for omitted triple-excitation terms. In Section~\ref{sec:systems}, we give some implementation details of our calculations for the nuclear ($^{16}$O and $^{22}$O) and electronic (6-electron parabolic quantum dots) systems considered. Results are presented in Section~\ref{sec:results}, and conclusions are presented in Section~\ref{sec:summary}. 

\section{\label{sec:formalism}Formalism}
In closed shell systems for which a single Slater determinant (SD) provides a reasonable reference state, any eigenstate of the $A$-body Hamiltonian may be written exactly as
\begin{equation}\label{eq:ex_state_expand}
|\Psi_\nu \rangle = C_0 | \Phi_0 \rangle +  \sum_{n=1}^A\frac{1}{n!}  \sum_{\substack{ i_1 \ldots i_n \\  a_1 \ldots a_n}} C^{a_1 \ldots a_n}_{i_1 \ldots i_n} |\Phi^{a_1 \ldots a_n}_{i_1 \ldots i_n} \rangle\,,
\end{equation}
where $| \Phi_0 \rangle$ is the reference SD, which we typically take as the Hartree-Fock approximation to the $A$-body ground state, and $|\Phi^{a_1 \ldots a_n}_{i_1 \ldots i_n} \rangle$ is the SD with the indicated number of particle-hole excitations out of the reference
\begin{equation}\label{eq:SD_def}
|\Phi^{a_1, \ldots a_n}_{i_1, \ldots i_n} \rangle = a^\dagger_{a_1} \cdots a^\dagger_{a_n} a_{i_n}  \cdots a_{i_1} |\Phi_0 \rangle\,.
\end{equation}
We use the convention for single-particle labels where $i,j,k,\ldots$ corresponds to occupied orbitals in the reference SD, $a,b,c,\ldots$ corresponds to unoccupied orbitals, and $q,r,s,\ldots$ is either.

In principle, an exact solution of the Schr{\"o}dinger equation in the complete SD basis would provide knowledge of the amplitudes $C_0, C^{a}_{i},C^{ab}_{ij},\ldots$ for each state $\nu$. In practice, for most systems the expansion must be severely truncated at some excitation rank $n \ll A$; one can then solve tractable generalized eigenvalue equations for the approximate amplitudes and energy eigenvalues. In this spirit, the Tamm-Dancoff approximation (TDA), the random phase approximation (RPA) and related equations-of-motion (EOM) techniques~\cite{Rowe:1968eq} offer computationally viable alternatives to full diagonalization for the calculation of excited states. Due to the necessary truncations, the types of correlations they capture can be limited significantly.  Consequently, simple methods such as these typically require the use of an effective Hamiltonian to account for the omitted degrees of freedom in order to give a quantitative description of spectra. In the following, we will show that the IMSRG is well-suited for this task, providing a convenient \emph{ab initio} framework to drive the Hamiltonian to a form where simple methods such as the TDA and EOM become very effective. 

\subsection{IMSRG}

We begin with a brief review of the IMSRG; for details, see the recent review articles~\cite{Hergert:2015awm,Hergert:2016etg}.
The similarity renormalization group (SRG) consists of a continuous sequence of unitary transformations that gradually suppress
off-diagonal matrix elements, driving the Hamiltonian towards a
band- or block-diagonal form~\cite{Glazek:1993il,Wegner:1994dk,Bogner:2006pc}.  Writing the transformed Hamiltonian as
\begin{align}
 \bar{H}(s)&=U(s)HU^\dagger(s)\equiv \bar{H}^d(s)+\bar{H}^{od}(s),\label{eq:ham_uni_trans}
\end{align}
where $\bar{H}^{d}(s)$ and $\bar{H}^{od}(s)$ are the arbitrarily defined diagonal and off-diagonal parts of the Hamiltonian. The evolution with the continuous flow parameter $s$
is given by 
\begin{align}
\label{eq:srg}
 \frac{d\bar{H}(s)}{ds}&=[\eta(s),\bar{H}(s)],
\end{align}
where $\eta(s)\equiv U(s)dU^{\dagger}(s)/ds$ is the (anti-Hermitian) generator of the transformation. Solving Eq.~\ref{eq:srg} with any generator amounts to performing the unitary transformation in Eq.~\ref{eq:ham_uni_trans} without explicitly constructing the $U(s)$ operator. The flexibility of the SRG stems from the fact that i) there are a multitude of choices one can make for $\eta(s)$ such that $\lim_{s\rightarrow\infty}\bar{H}^{od}(s)\rightarrow 0$, and ii) the partitioning of $\bar{H}(s)$ into diagonal and off-diagonal terms is completely arbitrary, allowing one to construct transformed Hamiltonians that are convenient for specific problems (e.g., ground-state versus excited-state calculations) through a suitable definition of $\bar{H}^{od}$.  

The ``in-medium'' part of the IMSRG refers to the use of normal-ordering with respect to an $A$-body reference state to capture the dominant effects of three- and higher-body interactions in a computationally efficient manner. Starting from the second-quantized Hamiltonian with two- and three-body interactions
\begin{align}
H &= \sum_{qr}T_{qr}a^{\dagger}_qa_r + \frac{1}{4}\sum_{qrst} V^{(2)}_{qrst}a^{\dagger}_qa^{\dagger}_ra_ta_s \nonumber\\
&+\frac{1}{36}\sum_{qrstuv}V^{(3)}_{qrstuv}a^{\dagger}_qa^{\dagger}_ra^{\dagger}_sa_{v}a_{u}a_{t}\,,
\end{align}
Wick's theorem can be used to normal-order $H$ with respect to an arbitrary $A$-body reference SD
\begin{align}\label{eq:no2b} 
H &= E_{ref} + \sum_{qr} f_{qr} \colon a^\dagger_q a_r \colon + \frac{1}{4} \sum_{qrst} \Gamma_{qrst} \colon a^\dagger_q a^\dagger_r a_t a_s \colon \notag \\
&+ \frac{1}{36}\sum_{qrstuv} W_{qrstuv} \colon a^\dagger_q a^\dagger_r a^\dagger_s a_v a_u a_t \colon\,.
\end{align}
Here, colons denote normal-ordered operator strings, whose expectation value in the reference state is zero by definition: $\langle \Phi_0 | \colon a^\dagger_q \cdots a_r \colon | \Phi_0 \rangle = 0$.  The key advantage of the normal-ordered representation is that the dominant mean-field contributions from three- and higher-body interactions are included in the $A$-dependent $0$-, $1$-, and $2$-body couplings $E_{ref}$, $f$, and $\Gamma$,
\begin{align}\label{eq:no_defs} 
E_{ref} &= \sum_i T_{ii}  + \frac{1}{2} \sum_{ij} V^{(2)}_{ijij} \notag\\ &
+ \frac{1}{6} \sum_{ijk} V^{(3)}_{ijkijk} \\
f_{qr} &= T_{qr} + \sum_i V^{(2)}_{qiri}  + \frac{1}{2}\sum_{ij} V^{(3)}_{qijrij}\\
\Gamma_{qrst} &= V^{(2)}_{qrst} + \sum_i V^{(3)}_{qristi} \\
W_{qrstuv} &= V^{(3)}_{qrstuv}\,, 
\end{align}
where antisymmetric two- and three-body matrix elements are assumed. 

Since the explicit inclusion of three-body interactions poses a significant challenge for most many-body methods, one common approach is to use the normal-ordered two-body approximation (NO2B), in which the $W_{qrstuv}$ matrix elements are neglected. In practice, the NO2B approximation has been shown to be an excellent approximation for a wide range of inter-nucleon interactions and nuclei~\cite{Hagen:2007zc,Roth:2012qf}. 

Applying Wick's theorem to evaluate Eq.~\ref{eq:srg} with $\bar{H}(s) = E_{ref}(s) + f(s)+ \Gamma(s)$ and $\eta(s) = \eta^{(1)}(s) + \eta^{(2)}(s)$
truncated to normal-ordered two-body operators, one obtains the
coupled IMSRG(2) flow equations~\cite{Tsukiyama:2011uq,Hergert:2015awm},
\begin{align}
  \totd{E_{ref}}{s}&= \sum_{ab}(n_a-n_b)\eta_{ab} f_{ba} 
    + \frac{1}{2} \sum_{abcd}\eta_{abcd}\Gamma_{cdab} n_a n_b\bar{n}_c\bar{n}_d
    \label{eq:imsrg2_m0b}\,,\\[5pt]
  \totd{f_{qr}}{s} &= 
  \sum_{a}(1+P_{qr})\eta_{qa}f_{ar} +\sum_{ab}(n_a-n_b)(\eta_{ab}\Gamma_{bqar}-f_{ab}\eta_{bqar}) \notag\\ 
  &\quad +\frac{1}{2}\sum_{abc}(n_an_b\bar{n}_c+\bar{n}_a\bar{n}_bn_c) (1+P_{qr})\eta_{cqab}\Gamma_{abcr}
 \label{eq:imsrg2_m1b}\,,\\[5pt]
  \totd{\Gamma_{qrst}}{s}&= 
  \sum_{a}(
    (1-P_{qr})(\eta_{qa}\Gamma_{arst}-f_{qa}\eta_{arst} ) \notag\\
  &-(1-P_{st})(\eta_{as}\Gamma_{qrat}-f_{as}\eta_{qrat} )) \notag\\
  &\quad+ \frac{1}{2}\sum_{ab}(1-n_a-n_b)(\eta_{qrab}\Gamma_{abst}-\Gamma_{qrab}\eta_{abst})
    \notag\\
  &\quad-\sum_{ab}(n_a-n_b) (1-P_{qr})(1-P_{st})\eta_{brat}\Gamma_{aqbs} 
    \label{eq:imsrg2_m2b}\,
\end{align}
where $P_{qr}$ is an operator that exchanges the indices $q$ and $r$, and $n_q$ are occupation numbers in
the reference state $|\Phi_0 \rangle$, with $\bar{n}_q\equiv 1-n_q$. Note that the $s$-dependence has been suppressed for brevity. 

In ground-state calculations for closed-shell systems, we seek to chose $\eta$ such that the transformation maps the correlated ground state to the reference state. In other words, the ground state of the transformed Hamiltonian is the reference state, and the eigenvalue (which is the zero-body piece of the transformed Hamiltonian) corresponds to the correlated ground-state energy. The generator $\eta$ is intimately tied to the ``off-diagonal'' terms in the Hamiltonian, which for the ground-state decoupling are
defined as those which couple the reference state to the Slater determinants with particle-hole excitations. In the NO2B approximation, this gives the following definition for the off-diagonal Hamiltonian, 
\begin{equation}\label{eq:gs_offdiag}
\bar{H}^{od} \in  \{ \Gamma_{{a}{b}{i}{j}},\Gamma_{{i}{j}{a}{b}},f_{ai},f_{ia}\} \,.
\end{equation} 
In the present work, we use the White generator~\cite{White:2002fk,Hergert:2015awm}, 
\begin{equation}\label{eq:white}
\eta^{(1)}_{ai} = f_{ai}/\Delta^a_i\hspace{0.5cm} \eta^{(2)}_{abij} = \Gamma_{abij}/\Delta^{ab}_{ij}\,,
\end{equation}
for ground-state-decoupling, along with the more numerically stable ``imaginary time'' generator~\cite{Hergert:2015awm}
for additional decoupling. The latter is defined as
\begin{equation}\label{eq:imtime}
\eta^{(1)}_{qr} = f^{od}_{qr}sgn(\Delta^q_r)\hspace{0.5cm} \eta^{(2)}_{qrst} = \Gamma^{od}_{qrst} sgn(\Delta^{qr}_{st})\,,
\end{equation}
where $\Delta$ are Epstein-Nesbet energy denominators~\cite{Hergert:2015awm}.
 
\begin{figure}[t]
  \centering
  \includegraphics[width=0.49\textwidth]{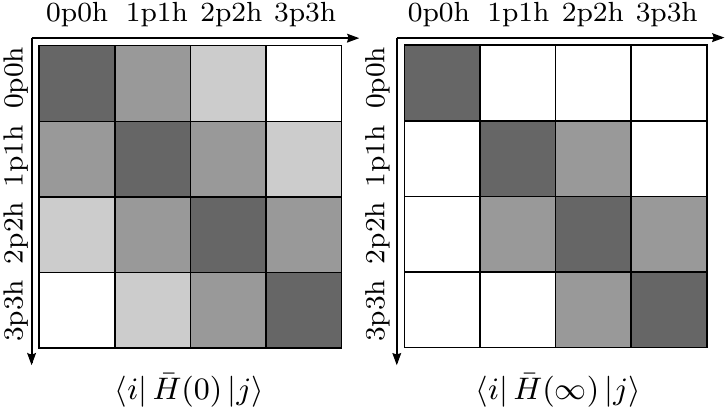} 
  \caption{\label{fig:schematic}Schematic representation of the initial and ground-state-decoupled Hamiltonians, $\bar{H}(0)$ and $\bar{H}(\infty)$, in the many-body Hilbert space spanned by particle-hole excitations of the reference state.}
\end{figure}

Integrating Eq.~\ref{eq:srg} with this generator drives $\bar{H}$ to a block diagonal form, where the reference state is decoupled from the $n$p$n$h excitation block as in Figure~\ref{fig:schematic}. The correlated ground-state energy is recovered for sufficiently large $s$ as the zero-body piece of the normal-ordered Hamiltonian
\begin{equation}\
E_{ref}(s \rightarrow \infty) \approx E_{gs}\,
\end{equation}  
up to IMSRG(2) truncation errors~\cite{Hergert:2015awm}. Numerical solutions of Eq.~\ref{eq:srg} are considered converged when the magnitude of the second order
perturbation theory contribution to the ground state energy is less than $10^{-6}$ MeV.  The IMSRG(2) has had great success in the description
of ground-state properties of closed-shell nuclei and electronic systems~\cite{Hergert:2015awm,sarahthesis}. 
In the following subsections we will develop an analogous IMSRG formalism for excited state calculations of varying sophistication. 

\subsection{\label{sec:sequential}Sequential Decoupling}
\subsubsection{TDA-IMSRG}
In the ground-state IMSRG, Hartree-Fock becomes exact for the ground state of the evolved Hamiltonian as $s\rightarrow\infty$.  It is natural to ask if an analogous decoupling can be designed so that some simple approximation for excited states becomes exact at the end of the evolution. To this end, we start with the well known Tamm-Dancoff approximation (TDA), where low-lying excited states are approximated as linear combinations of $1$p$1$h excitations of a reference Slater determinant~\cite{Ring:1980bb}, 
\begin{equation}\label{eq:tda_psi}
| \Psi_\nu^{TDA} \rangle = \sum_{ai} X^a_i a^\dagger_a a_i | \Phi_0 \rangle\,.
\end{equation} 
In this approximation, the Schr{\"o}dinger equation becomes:
\begin{equation}\label{eq:tda_eigval_eq}
\sum_{{b}{j}} ( f_{{a}{b}} \delta_{{i}{j}} - f_{{j}{i}} \delta_{{a}{b}} + \Gamma_{a{j}i{b}}) X^{{b}}_{{j}} = \omega^{TDA}_\nu X^{a}_{i}\,
\end{equation}
where $\omega^{TDA}_\nu = (E^{TDA}_\nu - E_{ref} )$.  For the case of a Hartree-Fock reference, the TDA is equivalent to diagonalizing $H$ on the subspace spanned by $|\Phi_0\rangle$ and the singly-excited $|\Phi^{a}_i\rangle$ Slater determinants. As it completely neglects ground-state correlations and higher-rank particle-hole excitations in the excited states, the TDA is deficient for Hamiltonians that exhibit significant coupling between the reference state and the higher particle-hole sectors, and between the $1$p$1$h and higher excitation blocks. The initial nuclear Hamiltonian (treated in the NO2B approximation) certainly falls into this class, as indicated by the left panel in Fig.~\ref{fig:schematic}.

On the other hand, we see that the ground-state-decoupled Hamiltonian in the right panel of Fig.~\ref{fig:schematic} is already in a semi-block-diagonal form with respect to particle-hole excitation rank.  Here, there are no correlations in the ground state since it's decoupled, and the $1$p$1$h block only couples directly to the $2$p$2$h block. Therefore, we expect that a TDA calculation for $\bar{H}(\infty)$ should be much more reliable than an analogous calculation for the initial Hamiltonian $\bar{H}(0)\equiv H$. 

In fact, the TDA becomes \emph{exact} for a subset of excited states -- modulo IMSRG(2) truncation errors -- if we apply a second IMSRG transformation that eliminates the coupling between different particle-hole excitation ranks, bringing the Hamiltonian to a block-diagonal form as in the left panel of Fig.~\ref{fig:tda_schematic}. Since these two transformations are applied sequentially, we label the ground-state decoupling as $U_1$, and the subsequent transformation to decouple the different particle-hole sectors as $U_2$. Therefore, we seek to construct
\begin{align}
\label{eq:H21}
\bar{H}_{21}(s)&\equiv U_2(s) \bar{H}_1(\infty)U^{\dagger}_2(s) \nonumber\\
&= U_2(s)U_1(\infty)HU^{\dagger}_1(\infty)U^{\dagger}_2(s)\,,
\end{align}
with the relevant off-diagonal terms for $U_2$ given in the NO2B approximation by
\begin{equation}\label{eq:Hod_u2} 
\bar{H}^{od}\in\{\Gamma_{i{c}{j}{k}}, \Gamma_{{b}{c}a{k}}\} + h.c.\,.
\end{equation}
Assuming the second IMSRG evolution converges, the transformed Hamiltonian becomes block-diagonal in particle-hole excitations
\begin{equation}\label{eq:U2_def} 
\langle \Phi^{a_1 \ldots a_n}_{i_1 \ldots i_n} | \bar{H}_{21}(\infty) | \Phi^{{a}'_1 \ldots {a}'_m}_{{i}'_1 \ldots {i}'_m} \rangle = 0  \hspace{0.5cm} ( n \neq m ) \,,
\end{equation} 
taking the schematic form shown in the left panel of Fig.~\ref{fig:tda_schematic}. Hereafter, we refer to this sequential decoupling as TDA-IMSRG, since the TDA becomes exact (up to IMSRG(2) truncation errors) when applied to $\bar{H}_{21}(\infty)$. 

\subsubsection{vTDA-IMSRG}

Due to the simple block-diagonal structure, the exact eigenvalues of $\bar{H}_{21}(\infty)$ can be obtained by diagonalizing each $n$p$n$h block separately, with the TDA being the simplest case. However, in actual calculations we find that the second $U_2$ transformation does not always converge with respect to the flow parameter $s$. Moreover, even when the $U_2$ evolution converges, the truncation errors due to the IMSRG(2) approximation can significantly degrade the unitary equivalence between the initial $H$ and $\bar{H}_{21}(\infty)$.  Heuristically, we expect that the loss of unitarity is due to the large number of off-diagonal terms driven to zero in the second transformation, which can lead to large induced three- and higher-body terms which are neglected in the IMSRG(2).

\begin{figure}[t]
  \centering
  \includegraphics[width=0.49\textwidth]{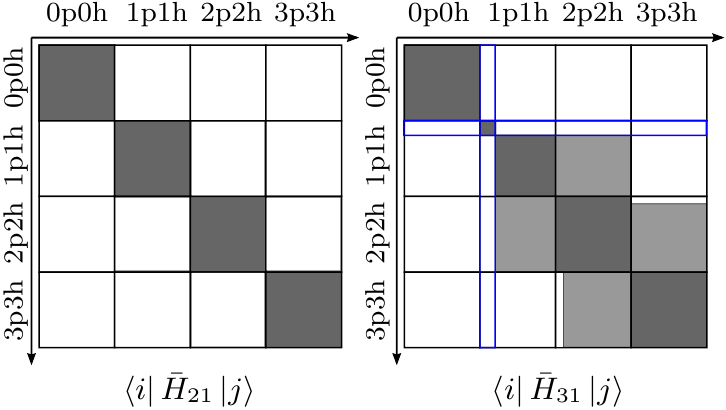} 
  \caption{\label{fig:tda_schematic}(Color online) Schematic representation of the sequentially-decoupled Hamiltonians in the many-body Hilbert space spanned by particle-hole excitations of the reference state. The left panel corresponds to the use of Eq.~\ref{eq:Hod_u2} where the entire $1$p$1$h sector is decoupled, and the right panel corresponds to Eq.~\ref{eq:vtda_offdiag} where just the valence $1$v$1$h excitations are decoupled. The latter corresponds to the small block in the upper left corner of the full $1$p$1$h-block. }  
\end{figure}

One way to minimize the loss of unitarity in the second transformation is to decouple a smaller portion of the $1$p$1$h configuration space in which the particle orbital is restricted to the lowest valence shell, as shown in the right panel of Fig.~\ref{fig:tda_schematic}. We refer to this as vTDA-IMSRG decoupling, and denote the new transformation by $U_3$ with 
\begin{align}
\bar{H}_{31}(s) &\equiv U_3(s) \bar{H}_1(\infty)U^{\dagger}_3(s) \nonumber\\
&= U_3(s)U_1(\infty)HU^{\dagger}_1(\infty)U^{\dagger}_3(s)\,.
\end{align}

To determine the form of $\bar{H}^{od}$ for the $U_3$ transformation, let us denote low-lying valence particle states as $a_v,b_v,c_v,\ldots$ and high-lying non-valence particle states as $a_q, b_q, c_q,\ldots$. If we don't distinguish between valence and non-valence particle states, we use the labels $a,b,c,\ldots$ as before. Performing TDA in the valence space alone modifies Eq.~\ref{eq:tda_psi} to

\begin{equation}\label{eq:tda_valence}
| \Psi_\nu^{VTDA} \rangle = \sum_{a_vi} X^{a_v}_i a^\dagger_{a_v} a_i | \Phi_0 \rangle\,,
\end{equation} 
and hence condition~\ref{eq:U2_def} is reduced to 
\begin{equation}\label{eq:coeffs2_svtda}
\langle \Phi^{{b}{c}}_{{j}{k}} | \bar{H}_{31} | \Phi_i^{a_v} \rangle = 0\,,
\end{equation} 
with the additional requirement
\begin{equation}\label{eq:coeffs2_svtda}
\langle \Phi^{a_q}_{{j}} | \bar{H}_{31} | \Phi_i^{a_v} \rangle = 0\,.
\end{equation} 
These two conditions are met if we choose 
\begin{equation}\label{eq:vtda_offdiag}
\bar{H}^{od}\in\{ \Gamma_{a_qi{j}a_v}, \Gamma_{i{c}{j}{k}}, \Gamma_{{b}{c}a_v{k}},f_{a_qa_v}\} + h.c.\,.
\end{equation} 
This definition of the off-diagonal terms 
is significantly reduced in scope from that of Eq.~\ref{eq:Hod_u2}, so we expect that the loss of unitarity caused by the IMSRG(2) truncation should be less severe.
The right panel of Figure~\ref{fig:tda_schematic} shows 
the schematic form of a successful vTDA-IMSRG(2) decoupling.  A 
vTDA-IMSRG(2) calculation will not leave the Hamiltonian block diagonal for all excitation ranks,
and will limit the number of states accessible to the calculation. However, if we are interested in 
only low-lying states, this calculation is much more stable than the full TDA-IMSRG(2). We note here that both TDA-IMSRG(2) and vTDA-IMSRG(2) are conceptually similar to the similarity-transformed EOM-CC method, reviewed recently in~\cite{Nooijen:2014}.  The TDA- and vTDA-IMSRG(2) evolution is considered converged when all excited states $\nu$ within the decoupled space obey
\begin{equation}\label{eq:ex_conv}
  | E_\nu(n) - E_\nu(n-1) | < \epsilon\,,
\end{equation}
where $n$ labels the timestep in $s$. In the present work, we take $\epsilon = 10^{-6}$ MeV.

\subsection{\label{sec:EOM}Equations-of-Motion Method}

The sequential decoupling is designed so that simple methods, such as TDA in the full $1$p$1$h space or TDA in the smaller $1$v$1$h  valence space, give exact eigenvalues of $\bar{H}_{21}(\infty)$ and $\bar{H}_{31}(\infty)$, respectively. However, both methods degrade the unitary equivalence to the original Hamiltonian due the second transformation in which a large number of matrix elements are driven to zero within the IMSRG(2) truncation.  While we anticipate that the loss of unitarity for the valence-space TDA should be less severe due to the ``gentler'' second transformation, the number of accessible excited states is much smaller due to the restricted configuration space. 

To avoid these limitations, we pursue a third strategy where we apply EOM techniques to approximately diagonalize the ground-state-decoupled Hamiltonian, eliminating the need for a second transformation.  Methods such as the TDA and RPA fall into a more general class 
of methods known as equations-of-motion (EOM) methods~\cite{Rowe:1968eq}. For any excited state, Eq.~\ref{eq:ex_state_expand} can be exactly rewritten in terms of 
a ladder-operator $X^\dagger_\nu$ and the correlated ground state
\begin{equation}\label{eq:EOM_assumption} 
|\Psi_\nu \rangle = X^\dagger_\nu | \Psi_0 \rangle\,.
\end{equation}
$X^\dagger_\nu$ is formally given by the dyad $|\Psi_{\nu}\rangle\langle\Psi_0|$, and can be written as a linear combination of 1- to A-body excitation and de-excitation operators. The energy eigenvalue problem can then be expressed in terms of the 
commutator of $H$ and $X^\dagger_\nu$:
\begin{equation}\label{eq:eom_eigval}
[H,X^\dagger_\nu]|\Psi_0 \rangle  = \omega_{\nu}X^\dagger_\nu|\Psi_0 \rangle\,,
\end{equation} 
where the excitation energy is $\omega_{\nu}=E_\nu - E_0$. Traditionally, the strength of EOM methods lies in the ability to
make controlled, computationally feasible approximations on the form of $X^\dagger_\nu$. Given some
approximation of the correlated ground state, the amplitudes of $X^\dagger_\nu$ can be
solved for in a generalized eigenvalue problem~\cite{Ring:1980bb}. In principle, the approximate ground
state can then be improved iteratively using the $X^\dagger_\nu$ and its Hermitian conjugate, which can then be used to get an improved $X^\dagger_\nu$, and so on.

One might naturally think to couple EOM methods with the IMSRG since
the reference state $|\Phi_0\rangle$ corresponds to the ground state of $\bar{H}_1\equiv U_1(\infty)HU^{\dagger}_1(\infty)$. Multiplying Eq.~\ref{eq:eom_eigval} by $U_1(\infty)$ and recalling that $U_1(\infty)|\Psi_0\rangle = |\Phi_0\rangle$ gives
\begin{equation}\label{eq:eom-imsrg_eigval}
 [\bar{H}_{1},\bar{X}^\dagger_\nu]|\Phi_0 \rangle  = \omega_{\nu} \bar{X}^\dagger_\nu|\Phi_0 \rangle\,,
\end{equation} 
where $\bar{X}^\dagger_\nu \equiv U_1(\infty)X^{\dagger}_{\nu}U^{\dagger}_1(\infty)$ only contains excitation operators since the reference state is annihilated by de-excitation operators.  We recover the TDA equations for the ground-state-decoupled Hamiltonian if we choose
\begin{equation}\label{eq:tda_ladder} 
\bar{X}^\dagger_\nu = \sum_{ai} \bar{X}^a_i a^\dagger_a a_i\,.
\end{equation} 
Alternatively, we may use a more sophisticated ladder operator which includes up to $2$p$2$h excitations,
\begin{equation}\label{eq:eom_ladder} 
\bar{X}^\dagger_\nu = \sum_{ai} \bar{X}^a_i a^\dagger_a a_i + \frac{1}{4}\sum_{a{b}i{j}} \bar{X}^{a{b}}_{i{j}} a^\dagger_a a^\dagger_{{b}} a_{{j}} a_i \,.
\end{equation}
Eq.~\ref{eq:eom_ladder} leads to a more complicated eigenvalue problem than the TDA, but it eliminates the need for a second transformation
as it includes a large portion of the correlations which are suppressed by $U_2$ or $U_3$.  Note that the EOM calculation with this ladder operator is
equivalent to diagonalizing $\bar{H}_1$ on the space of singly- and doubly-excited Slater determinants.

In general, the EOM ladder operator may have any excitation rank up to $A$p$A$h, 
which would constitute an exact diagonalization of $\bar{H}_1$.  Similarly, the level of truncation of the IMSRG equations can in principle be increased to the IMSRG($A$) level, where the unitary equivalence of $\bar{H}_1$ and $H$ is exact.  Therefore, EOM-IMSRG approximations are systematically improvable, allowing for 
EOM($m$)-IMSRG($n$) calculations, which will simply be referred to as EOM-IMSRG($m$,$n$). The calculations in the present work are carried out in the EOM-IMSRG($2$,$2$) approximation.

As a result of the vanishing de-excitation piece of $\bar{X}^\dagger_\nu$, Eq.~\ref{eq:eom-imsrg_eigval} has the advantage that it may 
be solved as a traditional eigenvalue problem using power-iteration methods such as the Lanczos algorithm. Such methods
only require knowledge of matrix-vector products. If $\bar{X}^\dagger_\nu$ is taken to be 
an eigenvector, the corresponding matrix-vector product is given by 
\begin{equation}\label{eq:mat_vec_prod}
[\bar{H}_1,\bar{X}^\dagger_\nu] = \{\bar{H}_1\bar{X}^\dagger_\nu\}_C 
\end{equation}
where the subscript $C$ denotes connected terms. 

An EOM-IMSRG(2,2) calculation proceeds as follows: 
\begin{enumerate}
\item Choose a reference state $|\Phi_0\rangle$.
\item Decouple the ground state via the IMSRG(2) with $\bar{H}^{od}$ defined as in Eq.~\ref{eq:gs_offdiag}.  
\item Solve Eq.~\ref{eq:eom-imsrg_eigval} using a ladder operator with single- and double-excitations, Eq.~\ref{eq:eom_ladder}.
\end{enumerate}

Note that ladder operators are spherical tensors of rank $J$ with definite parity, as they must connect 
the ground state to excited states of any desired spin $J^\pi$.  For this reason, EOM-IMSRG calculations are more
computationally demanding than both TDA-IMSRG and vTDA-IMSRG calculations. However, the relatively small rotation of the ground-state decoupling $U_1$ makes the EOM-IMSRG
equations far more numerically stable compared with both sequential decoupling approaches, which require a large secondary rotation $U_{2,3}$.  

\subsection{Perturbative triples correction}
\label{subsec:perttriples}

A straightforward correction to the EOM-IMSRG(2,2) spectra can be included via Rayleigh-Schr\"odinger perturbation theory that accounts for omitted triple-excitation terms 
in the ladder operator, Eq.~\ref{eq:eom_ladder}. The order zero wave function is taken to be the solution of the EOM-IMSRG(2,2), 
\begin{equation} 
|\tilde{\Psi}_\nu^{(0)}\rangle = |\bar{\Psi}_\nu\rangle = \bar{X}^{\dagger}_{\nu}|\Phi_0\rangle\,.
\end{equation}
Using Epstein-Nesbet partitioning of the Hamiltonian, the zero order energy is 
\begin{equation}
E_\nu^{(0)} =E_0 + \omega_\nu\,,
\end{equation} 
and the first order energy correction is zero by definition. The second order energy correction is then given by
\begin{equation}\label{eq:RSPT_2nd_order}
E_\nu^{(2)} = \langle \tilde{\Psi}_\nu^{(0)} | \bar{H}_1 \frac{\hat{Q}}{E^{(0)}_\nu - \bar{H}^{(0)}}  \bar{H}_1 | \tilde{\Psi}_\nu^{(0)} \rangle\,, 
\end{equation} 
where $\hat{Q}$ is the complement space projector
\begin{equation}\label{eq:Qspace} 
\hat{Q} = |\Phi_0\rangle \langle \Phi_0| + \sum_{\mu \neq \nu } |\bar{\Psi}_\mu \rangle \langle \bar{\Psi}_\mu | + \frac{1}{36} \sum_{ijkabc} | \Phi_{ijk}^{abc} \rangle \langle \Phi_{ijk}^{abc}| + \cdots 
\end{equation} 
Note that $\hat{Q} = 1-\hat{P}$, where $\hat{P} = |\bar{\Psi}_\nu \rangle \langle \bar{\Psi}_\nu | $ projects onto the particular solution of the EOM-IMSRG(2,2) for which we are calculating the perturbative correction. Since couplings between $|\bar{\Psi}_\nu\rangle$ and the reference state or $n$p$n$h excitations with $n\ge 4$ are zero in a ground-state-decoupled framework, and since couplings with $|\bar{\Psi}_{\mu \neq \nu} \rangle$ vanish due to the approximate diagonalization performed in the EOM-IMSRG(2,2) calculation, the triply-excited terms of Eq.~\ref{eq:Qspace} give the first non-vanishing contribution and Eq.~\ref{eq:RSPT_2nd_order} becomes
\begin{equation}\label{eq:pert_trips} 
 E^{(2)}_\nu = \frac{1}{36}\sum_{ijkabc}\frac{ \langle\Phi_0 | \bar{X}_\nu \bar{H}_1 |\Phi_{ijk}^{abc} \rangle \langle \Phi_{ijk}^{abc} |\bar{H}_1 \bar{X}^\dagger_\nu |\Phi_0 \rangle}{ E^{(0)}_\nu - \langle \Phi_{ijk}^{abc} | \bar{H}_1 | \Phi_{ijk}^{abc} \rangle  }\,.
\end{equation} 

Equation~\ref{eq:pert_trips} amounts to a perturbative energy correction that approximates the full EOM-IMSRG(3,2) energy, which is prohibitively expensive due to its $N_u^5 N_o^3$ scaling, where $N_o$ and $N_u$ refer to the number of occupied/unoccupied single-particle orbitals, and the need to store three-body matrix elements.  In practice, we write Eq.~\ref{eq:pert_trips} as: 
\begin{equation}
\label{eq:triples_correction}
E^{(2)}_\nu = \frac{1}{36}\sum_{ijkabc} \frac{|W_{abcijk}|^2}{D_{abc}^{ijk}} \\   
\end{equation} 
where
\begin{equation}
\label{eq:triples_denom}
D_{abc}^{ijk} = \omega_\nu -  \langle \Phi_{ijk}^{abc} | \bar{H} | \Phi_{ijk}^{abc} \rangle  \\
\end{equation}
and
\begin{equation}
\label{eq:triples_induced}
W_{abcijk} = [ \bar{H} , \bar{X}^\dagger_\nu ]_{abcijk} 
\end{equation} 
Storage of three-body matrix elements is not needed as Eqs.~\ref{eq:triples_correction}-\ref{eq:triples_induced} need only be calculated once for each excited state with manageable $N_u^4 N_o^3$ scaling. In the following, the inclusion of perturbative triples on top of EOM-IMSRG(2,2) will be referred to as the EOM-IMSRG(\{3\},2) approximation.

\section{ Systems and Interactions }
\label{sec:systems}
Before presenting the results of our calculations of excited states in 2d parabolic quantum dots and $^{16,22}$O nuclei, we present some details of our implementations for both systems.

\subsection{ Quantum Dots }  
Parabolic quantum dots consist of $A$ electrons confined by a harmonic oscillator potential in two dimensions. In atomic units, the Hamiltonian is given by:
\begin{equation}\label{eq:qd_hamiltonian}
H = \sum_{i=1}^A [\frac{1}{2} p_i^2 + \frac{1}{2} \omega^2 r_i^2] + \frac{1}{2} \sum_{i \ne j }^A \frac{1}{|{\bf r}_i - {\bf r}_j |} \,.
\end{equation}  
Quantum dots provide an excellent testing ground for approximate many-body methods, as the strength of many-body correlations can be controlled by varying $\omega$, with smaller values corresponding to stronger correlations, and comparisons can be made to exact full configuration interaction (FCI) calculations in sufficiently small bases. In the present work, all calculations are performed for the 6-electron system.
\begin{figure}[h]
  \centering
  \includegraphics[width=0.45\textwidth]{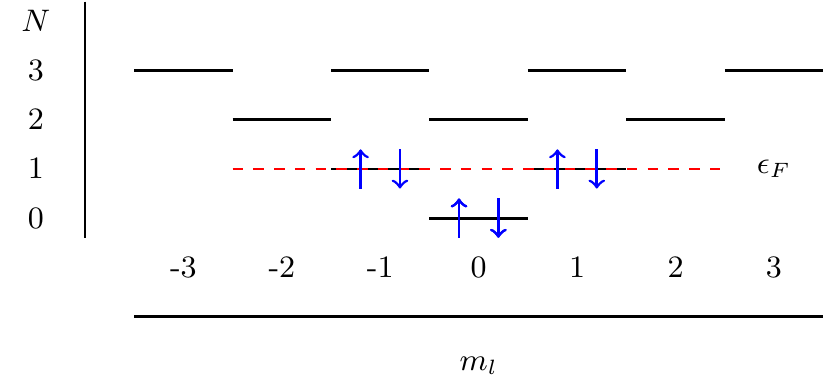} 
  \caption{\label{fig:qd_schematic}(Color online) Orbital scheme for a 6-electron two-dimensional quantum dot in a model space consisting of four major shells. Each orbital has a distinct orbital angular momentum projection $m_l$ and spin-projection $m_s$. }
\end{figure}

Figure~\ref{fig:qd_schematic} depicts the orbital scheme used to define the single-particle basis. The orbitals for this system are the solutions to the one body piece of Eq.~\ref{eq:qd_hamiltonian}, i.e. solutions of 
the two-dimensional isotropic harmonic oscillator problem, and are characterized by the projection of 
orbital momentum and spin quantum numbers, given by $m_l$ and $m_s$ respectively, and the usual principal quantum number $n$. The single-particle energy of a given orbital is $E_N = \omega(N+1)$, where $N = 2n + |m_l|$.  For FCI calculations, the full configuration space is composed of all possible A-body Slater determinants within the chosen single-particle basis. In this work, we truncate the single-particle basis to orbitals $N\leq3$, as we are only concerned with having exact FCI results to compare against for our approximate vTDA-IMSRG(2) and EOM-IMSRG(2,2) calculations.  For the latter two methods, the Hartree-Fock equations are solved by expanding the unknown HF orbitals in the truncated $N=3$ oscillator basis. The Hamiltonian in Eq.~\ref{eq:qd_hamiltonian} is then written in the self-consistent HF basis and normal-ordered with respect to the ground-state Slater determinant for the $A=6$ case. Finally, the in-medium 0-,1-, and 2-body pieces of the normal-ordered Hamiltonian are used as initial conditions for the numerical solution of the relevant IMSRG(2) equations. 

\subsection{ Finite Nuclei } 
In our calculations of the $^{16,22}$O nuclei, we start with the intrinsic $A$-body Hamiltonian
\begin{equation}\label{eq:nuclear_int_hamiltonian} 
H = (1-\frac{1}{A})T^{(1)} + T^{(2)} + V^{(2)}\,, 
\end{equation} 
where the center-of-mass (COM) kinetic energy is subtracted out, leading to a two-body kinetic energy piece $T^{(2)}$ as well as the $A$ dependence of the one-body term, see Ref.~\cite{Hergert:2015awm}.  Since the primary purpose of the present work is to perform proof-of-principle calculations with the EOM-IMSRG(2,2) method, and not to make detailed comparisons to experiment, we neglect three-nucleon interactions for simplicity and consider the N$^3$LO (500 MeV) input nucleon-nucleon potential of Entem and Machleidt (EM)\cite{Entem:2003th}, softened by free-space SRG evolution at the two-body level to two different resolution scales, $\lambda = 2.0$ fm$^{-1}$ and $\lambda=3.0$ fm$^{-1}$.  As with the quantum dot calculations, the Hartree-Fock equations are first solved by expanding the unknown HF orbitals in a spherical harmonic oscillator basis truncated to oscillator states obeying $2n+l \leq e_{\rm max}$, where $e_{\rm max}$ is sufficiently large so that the results are approximately independent of the $\hbar\omega$ value of the underlying oscillator basis. Once a converged solution is obtained, the Hamiltonian is normal-ordered with respect to the ground-state Slater determinant, and the resulting in-medium zero-, one-, and two-body operators supply the initial values for the IMSRG(2) flow equations. 

Since nuclei are self-bound objects governed by a translationally-invariant Hamiltonian, an exact solution of the Schr\"odinger equation must factorize into the product of a wave function for the physically relevant intrinsic motion times a wave function for the COM coordinate,
\begin{equation}
\label{eq:comfac}
|\Psi\rangle = |\psi\rangle_{\rm in}\otimes|\psi\rangle_{\rm cm}\,.
\end{equation}

There are two strategies to rigorously guarantee this factorization; one can work in a translationally-invariant basis from the outset, or one can work in a so-called full $N\hbar\omega$ model space comprised of all $A$-particle harmonic oscillator Slater determinants with excitations up to and including $N\hbar\omega$. Neither choice is optimal since the former is limited to light nuclei due to the factorial scaling of the required antisymmetrization, while the latter limits the choice of the single-particle orbitals to the harmonic oscillator basis and doesn't carry over to methods such as coupled cluster theory and the IMSRG where it is more natural to define the model space via an energy cutoff (e.g., $2n+l \leq e_{\rm max}$) on the single-particle states. In the case of calculations with an $e_{\rm max}$ cutoff, there is no analytical guarantee that the COM and intrinsic wave functions factorize.

In Ref.~\cite{Hagen:2009fk}, Hagen and collaborators gave an ingenious prescription to diagnose whether or not Eq.~\ref{eq:comfac} is satisfied in such calculations. The basic idea is to assume that the factorized COM wave function is a Gaussian, and is therefore the ground state with eigenvalue zero of the shifted COM Hamiltonian $H_{\rm cm}(\tilde{\omega})$,
\begin{equation}
H_{\rm cm}(\tilde{\omega})= \frac{{\bf P}^2}{2 m A }  + \frac{1}{2} m A \tilde{\omega}^2 {\bf R}^2 - \frac{3}{2}\hbar\tilde{\omega}\,,
\end{equation}
where $m$ is the average nucleon mass, and ${\bf P}$ and ${\bf R}$ are the center-of-mass momentum and position operators, respectively.
Note that $\tilde{\omega}\neq \omega$ in general, where $\omega$ is the frequency of the underlying oscillator basis. The prescription to find $\tilde{\omega}$ involves solving a quadratic equation

\begin{equation}
\label{eq:omegatilde}
\hbar\tilde{\omega} = \hbar\omega +\frac{2}{3}E_{\rm cm}(\omega) \pm \sqrt{\frac{4}{9}(E_{\rm cm}(\omega))^2 +\frac{4}{3}\hbar\omega E_{\rm cm}(\omega)},
\end{equation}
where
\begin{align}
E_{\rm cm}(\omega)&\equiv \langle\Psi|H_{\rm cm}(\omega)|\Psi\rangle \\
&=\lim_{s\rightarrow\infty}\langle\Phi|\bar{H}_{\rm cm}(\omega;s)|\Phi\rangle\,,
\end{align}
and $\bar{H}_{\rm cm}(\omega;s)$ is the consistently evolved COM operator. Since there are two roots of Eq.~\ref{eq:omegatilde}, we choose the one that gives a smaller value for $E_{\rm cm}(\tilde{\omega})$ in the ground-state IMSRG calculation. Once $\tilde{\omega}$ is obtained from the ground-state calculation, we follow the Lawson-Gloeckner prescription~\cite{Gloeckner:1974gb} for the EOM-IMSRG calculations by adding the following term to the bare intrinsic Hamiltonian 
\begin{equation} 
H_{L} = \beta H_{{c}.\!{m}.\!}(\tilde{\omega}),
\end{equation} 
and performing the IMSRG(2) ground-state evolution and the subsequent EOM calculation. Note that $\beta$ is an adjustable parameter used to shift spurious COM excitations in the EOM-IMSRG spectrum to irrelevantly high energies.  

\begin{figure}[b]
  \centering
  \includegraphics[width=0.49\textwidth]{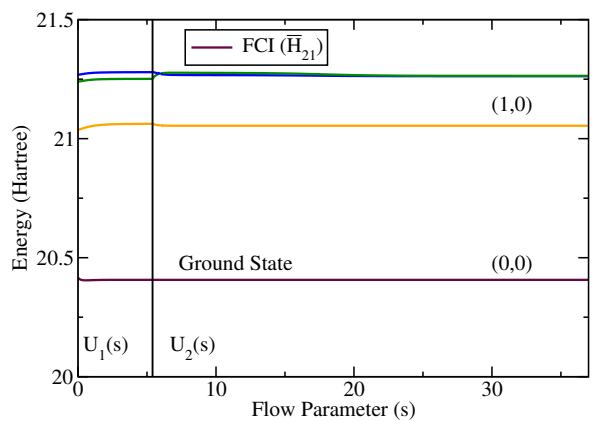} 
  \caption{\label{fig:fci_flow_full}(Color online) Ground state and low-lying $(M_L,M_S)=(1,0)$ excited states of a 6-electron quantum dot in an $\omega =1.0$ trap. The FCI calculations are performed using the flowing TDA-IMSRG(2) Hamiltonian in an $N=3$ model space.}
\end{figure}

\section{Results}
\label{sec:results}
\subsection{ Full Configuration Interaction Analysis of TDA-IMSRG(2) and vTDA-IMSRG(2)}
The sequential TDA-IMSRG decoupling discussed in in Section~\ref{sec:sequential} is designed so that a $1$p$1$h configuration interaction calculation is exact for a subset of eigenvalues. If the IMSRG evolution is carried out without truncation, the transformation is unitary and the eigenvalues of the evolving Hamiltonian are invariant throughout the flow. However, since both stages of the decoupling are carried out in the IMSRG(2) approximation, exact unitary equivalence with the initial Hamiltonian is lost due to the neglect of induced three- and higher-body terms.  One way to assess the loss of unitarity is to perform FCI calculations using the evolved Hamiltonian at different values of the flow parameter. If no truncations are made, then the transformation is exactly unitary and the FCI spectra are $s$-independent. Therefore, the degree of $s$-dependence in the spectra provides a measure of the truncation errors associated with the IMSRG(2) approximation.

Figure~\ref{fig:fci_flow_full} demonstrates the behavior of FCI as a function of $s$ for a few low-lying energy levels of the 6-electron quantum
dot system, with the single-particle model space truncated to the first four oscillator shells. The FCI calculations are performed using the evolved Hamiltonian at intermediate steps in the sequential decoupling defined in Eq.~\ref{eq:H21}, where the first stage of the IMSRG evolution decouples the ground state, and the second stage decouples the particle-hole excitations. The vertical black line in Fig.~\ref{fig:fci_flow_full} indicates the endpoint of the ground-state decoupling $U_1$, and the beginning 
of the secondary $1$p$1$h decoupling $U_2$.  The ground state and first $(M_L,M_S)=(1,0)$ excited state are only weakly dependent on $s$ for both transformations, indicating that the loss of unitarity from the IMSRG(2) is small.  

\begin{figure}[t]
  \centering
  \includegraphics[width=0.49\textwidth]{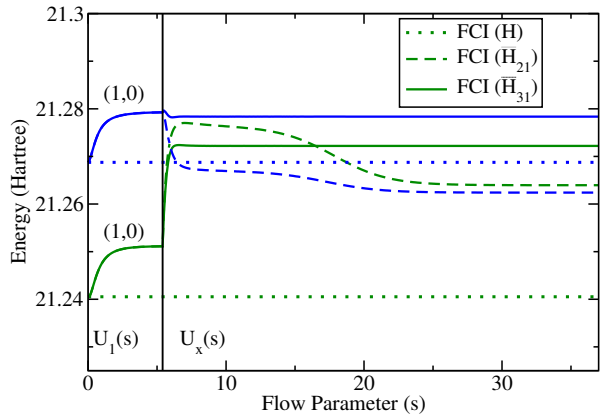} 
  \caption{\label{fig:fci_flow}(Color online) $(M_L,M_S)=(1,0)$ excited states
  of a 6-electron quantum dot in an $\omega =1.0$ trap with an $N=3$ model space. Three FCI calculations are shown; two using the flowing TDA-IMSRG(2) and vTDA-IMSRG(2) Hamiltonians corresponding to each choice of the secondary decoupling $U_{x}(s)$ (where $x=2$ or $3$), and one using the bare Hamiltonian. }
\end{figure} 

If we zoom in on the second and third excited states in Fig.~\ref{fig:fci_flow}, however, we see a more pathological behavior. For reference, the FCI results using the initial Hamiltonian are indicated by the horizontal dotted lines. In the second stage of the transformation (i.e., to the right of the vertical line), dashed lines show FCI results for the flowing $\bar{H}_{21}(s)$ during the TDA-IMSRG(2) decoupling, while solid lines show FCI performed for the flowing $\bar{H}_{31}(s)$ during the vTDA-IMSRG(2) decoupling. Here, the valence space includes all $1$p$1$h excitations into the $N=2$ shell. Apart from a small region in $s$ after the second $U_3$ evolution is initiated, the $\bar{H}_{31}(s)$ FCI spectra show minimal $s$-dependence, and the calculation has numerically converged at $s\approx 11.5$.  On the other hand, the $\bar{H}_{21}(s)$ spectra behave erratically; the two levels cross and exhibit significant $s$-dependence during most of the $U_2$ evolution. This is consistent with our naive expectations that the IMSRG(2) truncation errors should be smaller for the valence $1$p$1$h decoupling since fewer matrix elements are being driven to zero.  While this does not conclusively prove that induced many-body terms are always less problematic for the $U_3$ evolution, we note that in calculations of larger systems, the TDA-IMSRG(2) decoupling often leads to non-convergent energies and numerical instabilities, which again is consistent with our intuition that induced many-body interactions are more problematic for the ``stronger'' $U_2$ transformation. Until these instabilities are better understood, it appears the vTDA-IMSRG(2) is the preferred sequential decoupling method to target excited states.

\begin{figure}[t]
  \centering
  \includegraphics[width=0.49\textwidth]{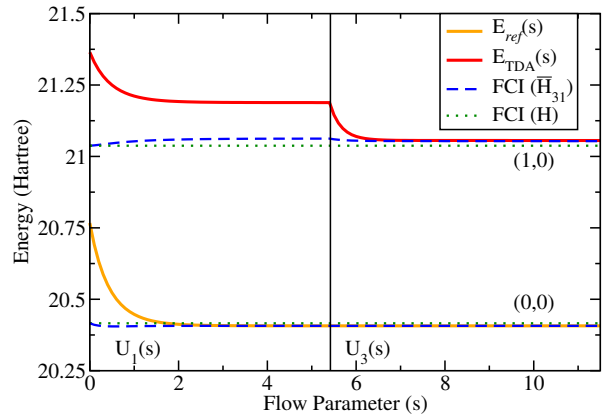} 
  \caption{\label{fig:tda_flow}(Color online) FCI and TDA calculations of the first $(M_L,M_S)=(1,0)$ excited state using the flowing vTDA-IMSRG(2) Hamiltonian for a 6-electron quantum dot with $N = 3$ and $\omega = 1.0$. For reference, the flowing $0$-body part of the Hamiltonian $E_{ref}(s)$ and the FCI results for the bare Hamiltonian are also shown.} 
\end{figure} 
\begin{figure*}[t!]
  \centering
   \includegraphics[width=0.49\textwidth]
 {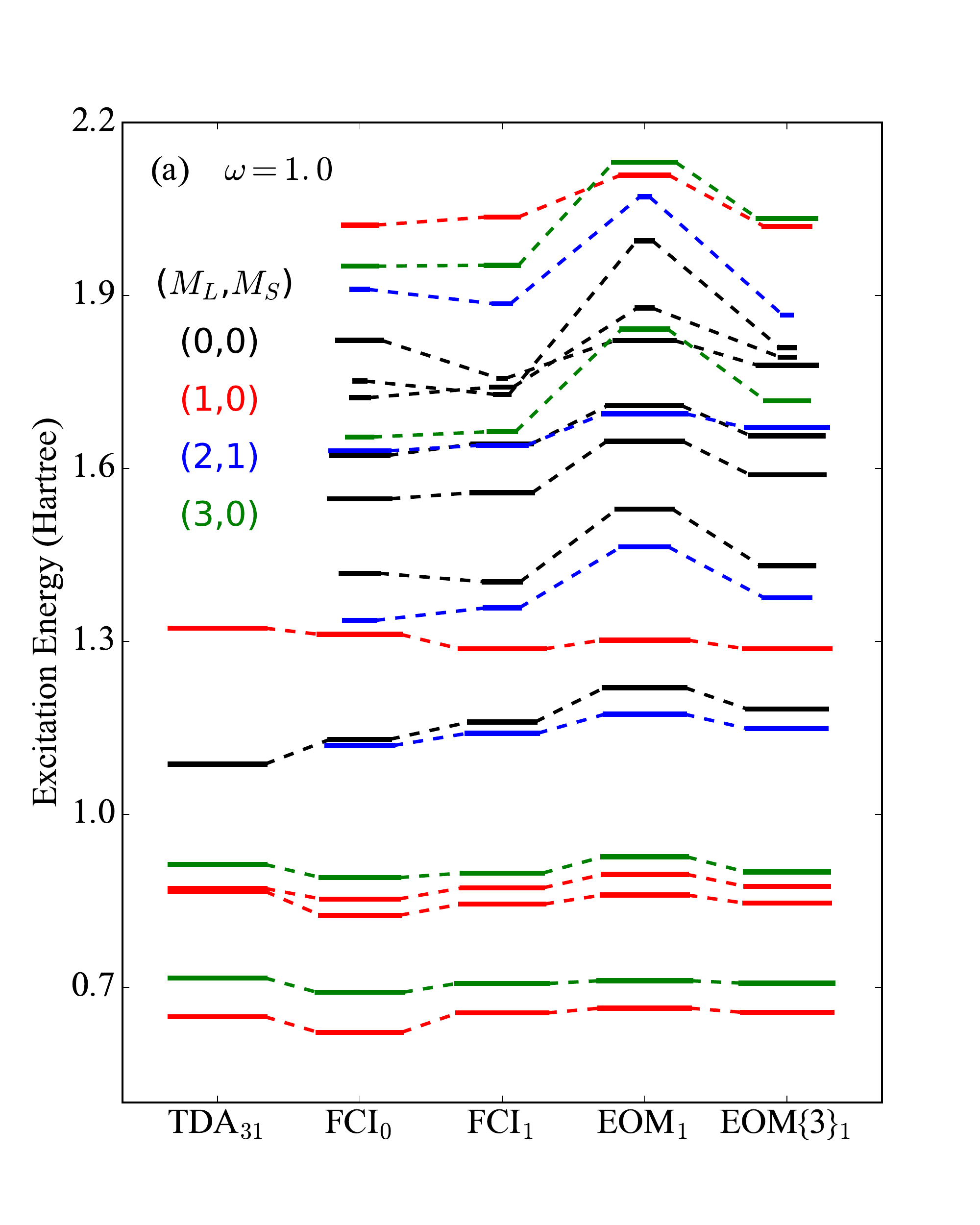}~~~%
 \includegraphics[width=0.49\textwidth]%
 {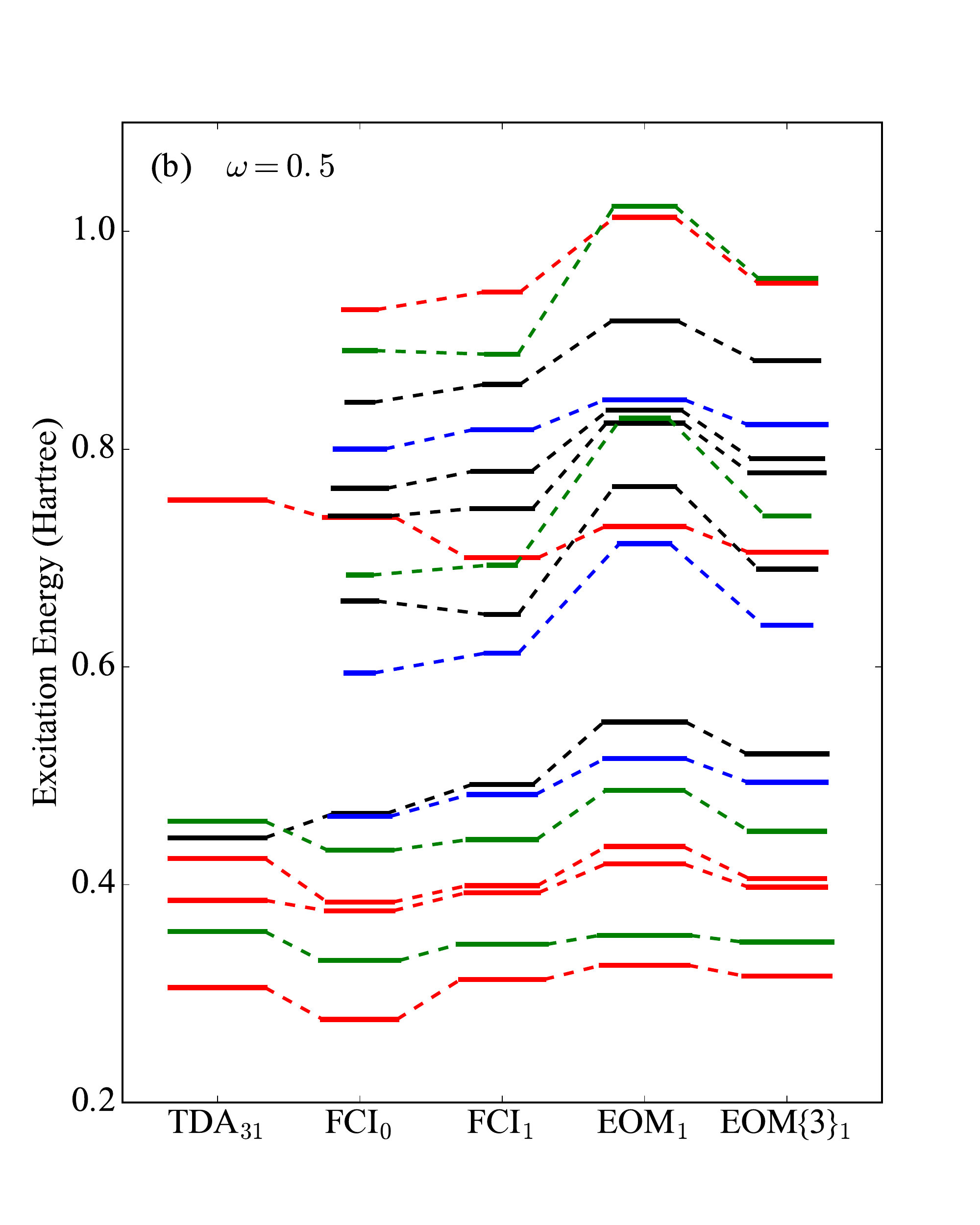}
   \caption{\label{fig:spectrum}(Color online) Selected excitation spectra of 6-electron quantum dots for $\omega = 1.0$ (a) and $\omega=0.5$ (b) performed in an $N=3$ single-particle basis. The quantum numbers of the various states are color-coded as ($M_L$,$M_S$) = {\color{red} (0,0)},{\color{black} (1,0)},{\color{blue} (2,1)},{\color{FXGreen} (3,0)}. The calculated spectra are displayed for five
different many-body approaches, where the subscript indicates which Hamiltonian the respective method is applied to. For example FCI$_0$ and FCI$_1$ denote FCI calculations on the bare and ground-state-decoupled Hamiltonians respectively, TDA$_{31}$ denotes a TDA calculation on the vTDA-decoupled Hamiltonian, etc. The lengths of the plotted energy levels indicate the $1$p$1$h content of the state as defined in Eqs.~\ref{eq:FCI_crit}-\ref{eq:TDA_crit}. }
\end{figure*}

Figure~\ref{fig:tda_flow} demonstrates the utility of the vTDA-IMSRG(2) sequential decoupling in TDA calculations for the same 6-electron quantum dot. 
For comparison, flowing and bare FCI calculations for the ground state and first excited state are included, in addition to the flowing $0$-body component of the Hamiltonian $E_{ref}(s)$.  At $s=0$, $E_{ref}(0)$ and $E_{TDA}(0)$ are rather poor approximations to the exact ground and excited state energies, since they correspond to the Hartree-Fock and Tamm-Dancoff approximations calculated with the bare Hamiltonian.  As $s$ increases, more and more many-body correlations are resummed into the flowing Hamiltonian, and the corresponding $E_{ref}(s)$ and $E_{TDA}(s)$ become better approximations to the exact results. By $s\approx 5.4$, the ground-state
decoupling is complete and $E_{ref}(s\approx5.4)$ is in excellent agreement with the exact ground state, while $E_{TDA}(s\approx 5.4)$ gives a significantly improved estimate of the excited state.  As the secondary decoupling progresses, $E_{TDA}(s)$
approaches the flowing FCI calculation, and is a very good approximation of the exact bare FCI calculation. Note that the ground-state
energy is unaffected by $U_3$ as the reference state remains decoupled from the higher particle-hole sectors throughout. The diagonalization of the valence $1$p$1$h block is much less computationally demanding than the full Hamiltonian matrix of FCI, despite similar quality of results. Despite the success of the TDA-IMSRG(2) method, it produces useful results only for states which have exceedingly strong $1$p$1$h character. Restricting 
ourselves to a valence space limits the number of accessible states even further. Therefore, we turn now to the more versatile EOM-IMSRG. 

\subsection{Quantum Dots Energy Spectra using EOM-IMSRG and TDA-IMSRG}   
Figure~\ref{fig:spectrum} shows vTDA-IMSRG(2), EOM-IMSRG(2,2) and EOM-IMSRG(\{3\},2) (labeled TDA$_{31}$, EOM$_1$, EOM\{3\}$_1$, respectively) spectra for two different quantum dots, along with FCI calculations performed for the bare Hamiltonian (FCI$_0$) and FCI calculations using the ground-state-decoupled Hamiltonian $\bar{H}_1$  (FCI$_1$). The length of the lines indicate the $1$p$1$h content of a given state which we define as
\begin{align}
\label{eq:FCI_crit} 
n(1p1h)_{FCI_0} &= \sum_{ph} |C_h^p|^2   \\ 
\label{eq:FCI1_crit} 
n(1p1h)_{FCI_1} &= \sum_{ph} |{(\bar{C}_1)}_h^p|^2 \\
 \label{eq:EOM_crit} 
n(1p1h)_{EOM_1} &= \sum_{ph} |{(\bar{X}_1)}^p_h|^2 \\
\label{eq:TDA_crit} 
n(1p1h)_{TDA_{31}} &= \sum_{vh} |{(\bar{X}_{31})}^v_h|^2 = 1\,.
\end{align}  
Note that  this quantity is defined differently depending on the particular unitary transformation, so a direct 1-to-1 comparison can be misleading. For instance, in the vTDA-IMSRG calculations, the excited states are completely $1$p$1$h in the unitarily-transformed frame, hence all of the lines in the TDA$_{31}$ column are the maximum possible length. Since FCI$_1$ and EOM$_1$ are performed for the same operator,  $\bar{H}_1$, a direct comparison of Eq.~\ref{eq:EOM_crit} and \ref{eq:FCI1_crit} is consistent. We note that EOM-IMSRG(\{3\},2) partial norms are corrected by normalizing with the wave function corrected to first order in perturbation theory, resulting in a slight decrease. 

In Figure~\ref{fig:spectrum} we show four sets of states with the indicated quantum numbers chosen to demonstrate the robustness of the EOM-IMSRG method. For odd parity states such as ($M_L$,$M_S$) = (1,0) and (3,0), we see that those which are strongly of $1$p$1$h nature are well-described by both vTDA-IMSRG and EOM-IMSRG methods. We also note that the EOM-IMSRG reproduces the FCI$_1$ partial norms nicely for these states, indicating that the EOM-IMSRG(2,2) is a good approximation to the full diagonalization of $\bar{H}_1$.  The EOM-IMSRG spectra degrades somewhat for even parity states, since the sizable shell gap at the Fermi level tends to suppress the $1$p$1$h dominance for such states,  and at higher excitation energies. However, it bears repeating that the EOM-IMSRG is significantly more flexible than the vTDA-IMSRG, as the latter is intrinsically unable to access even parity and/or higher excited states without expanding the model space to include the entire $1$p$1$h configuration space, which often leads to numerical instabilities and/or erratic convergence.

Another advantage of the EOM-IMSRG approach is that it can be systematically improved. EOM-IMSRG(\{3\},2) corrections significantly reduce the errors in the EOM-IMSRG(2,2) approximation at a manageable computational cost. Excitation energies, which are
consistently over-estimated by the EOM-IMSRG(2,2) calculation, are consistently reduced by the perturbative triples correction, bringing results into better agreement with the FCI$_{1}$ and FCI$_0$ spectra. The quality of EOM-IMSRG(\{3\},2) energies is still dependent on higher excitation rank content, but $1$p$1$h and $2$p$2$h states are described well 
in this approximation. 

\begin{figure}[t]
  \centering
  \includegraphics[width=0.49\textwidth]{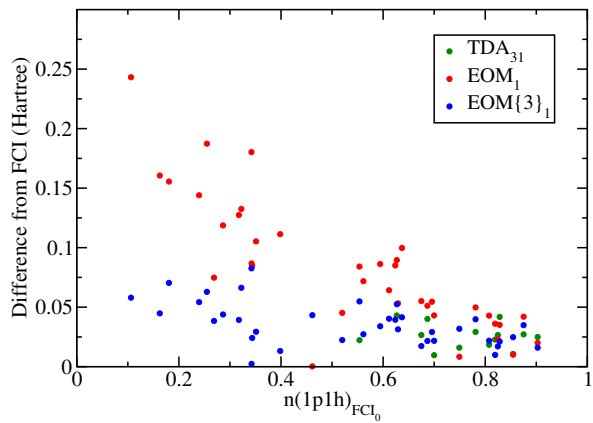} 
  \caption{\label{fig:errors}(Color online) Absolute difference between quantum dot excitation energies
  calculated via IMSRG methods and those calculated with FCI on the bare Hamiltonian. Each point
  corresponds to an EOM or TDA energy level in Figure \ref{fig:spectrum}. }
\end{figure} 

The quality of IMSRG results degrades as the importance of correlation in the system increases. This is seen clearly in the right panel of Figure~\ref{fig:spectrum} for the smaller trap frequency $\omega=0.5$. Nevertheless, the perturbative triples correction still gives substantial improvement. One can easily spot a correlation between the errors of either method and the bare FCI $1$p$1$h amplitudes. Figure~\ref{fig:errors} shows 
the absolute difference between the FCI$_0$ excitation energy and those calculated via EOM-IMSRG(2,2) or TDA-IMSRG(2), plotted against the bare FCI partial norm for each state. A clear inverse proportional relationship can be seen. Accessible TDA-IMSRG(2) results are generally more accurate than EOM-IMSRG(2,2). This is expected, as a successful TDA-IMSRG calculation should fully decouple the relevant excited states from truncated terms in the configuration expansion, where EOM-IMSRG(2,2) ignores some non-zero couplings by definition. This difference is for the most part erased by the EOM-IMSRG(\{3\},2) triples correction. 
The root-mean square deviations from the FCI results are $0.095$ Hartree for EOM-IMSRG(2,2) and $0.031$ Hartree for EOM-IMSRG(\{3\},2).

\begin{figure}[t]
  \centering
  \includegraphics[width=0.49\textwidth]{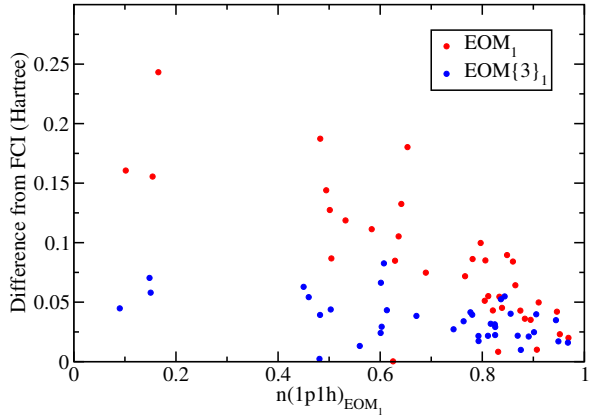} 
  \caption{\label{fig:errors_EOM}(Color online) Same as figure~\ref{fig:errors}, except energies are plotted with respect to EOM-IMSRG(2,2) partial norms.}
\end{figure} 
In larger spaces, FCI calculations are not feasible, so we should also consider the relationship between the error   
and the EOM$_1$ partial norm of Eq.~\ref{eq:EOM_crit}. Figure~\ref{fig:errors_EOM} demonstrates this relationship, using the same states displayed in Figure~\ref{fig:errors}. It is evident that EOM-IMSRG overestimates the $1$p$1$h content of calculated states, however there is a modestly linear relationship 
between the error and the EOM partial norm. This is a useful tool to gauge the reliability of EOM-IMSRG calculations in larger spaces, as EOM
amplitudes are immediately available after solution of Eq.~\ref{eq:eom_eigval}. 

\subsection{Results in Nuclear Physics} 
\begin{figure}[t]
  \centering
  \includegraphics[width=0.49\textwidth]{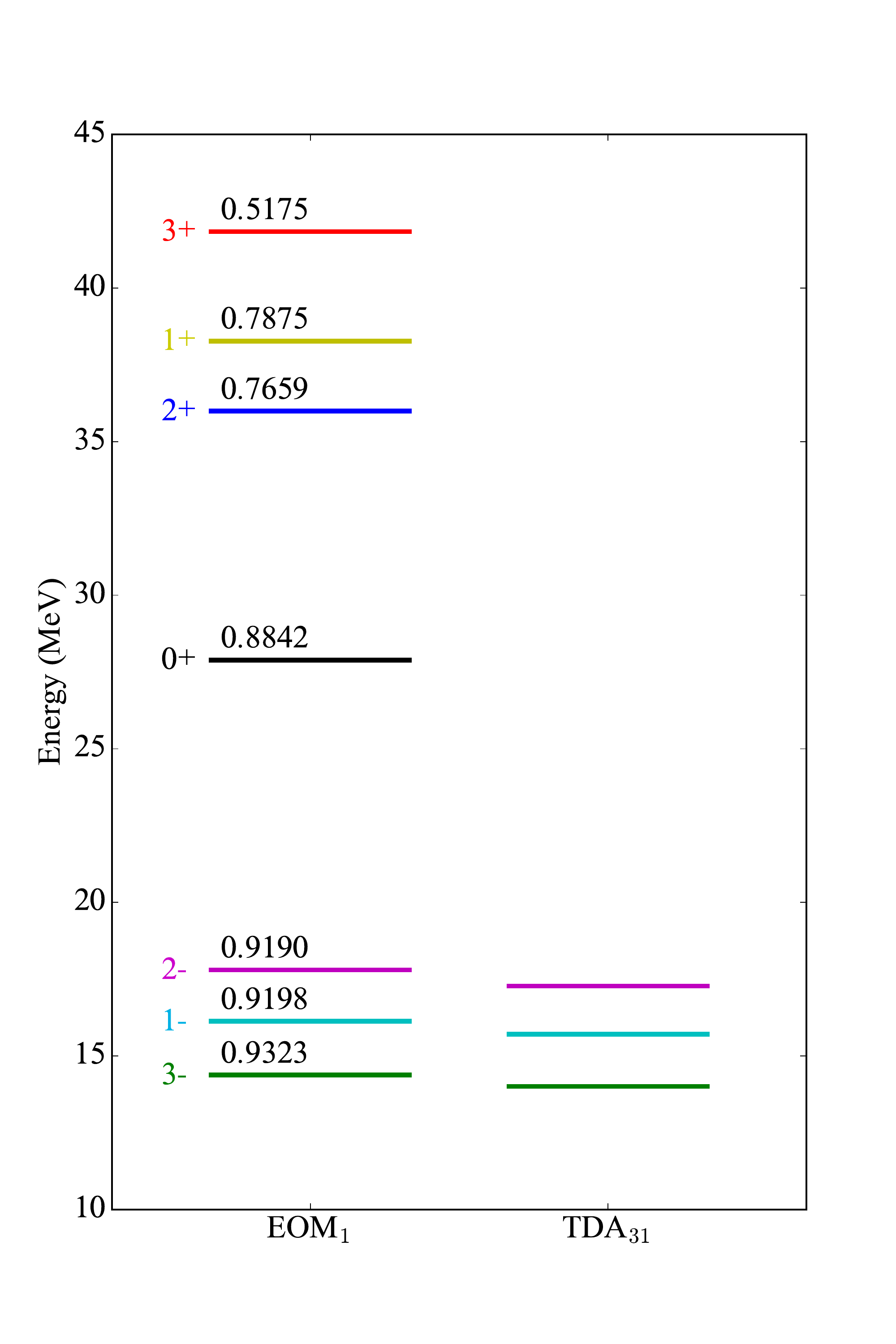} 
  \caption{\label{fig:TDA_v_EOM}(Color online) Lowest $^{16}$O excitation energies plotted for various quantum numbers, calculated with EOM-IMSRG(2,2) and vTDA-IMSRG(2) starting from the N$^3$LO (500 MeV) NN interaction of Entem and Machleidt (EM)~\cite{Entem:2003th}, softened by free-space SRG evolution to $\lambda = 2.0$ fm$^{-1}$. The single-particle basis is given by $\hbar \omega$ = 24.0 MeV and $e_{max}=8$. Above each plotted energy level from the EOM-IMSRG(2,2) calculation is the $1$p$1$h partial norm of Eq.~\ref{eq:EOM_crit}.} 
\end{figure}

Applying both the vTDA-IMSRG(2) and EOM-IMSRG(2,2) to finite nuclei, we find a clear preference for the latter method. Unfortunately, the promising results of vTDA-IMSRG(2) calculations in quantum dots do not carry over to nuclei, as uncontrolled numerical instabilities in the secondary transformation render the vTDA-IMSRG unusable for systems with strong correlations. Until these instabilities are better understood and overcome, sequential decoupling appears to be appropriate only for computations in doubly-magic nuclei.  Figure~\ref{fig:TDA_v_EOM} depicts the lowest excitation energies of $^{16}$O calculated at several different angular momenta and parities.   
We find that the vTDA-IMSRG(2) tracks well with EOM-IMSRG(2,2) for low-lying $1$p$1$h dominant states, but is non convergent for all others. In the left column, the $1$p$1$h partial norm (Eq.~\ref{eq:EOM_crit}) of the EOM wave function is listed above the corresponding energy for that state.  While the $0^+$ state has strong $1$p$1$h content in the EOM calculation, the vTDA-IMSRG(2) fails to converge beyond the three lowest excited states\footnote{We are not attaching any physical meaning to states at such unphysical high excitation energies. Rather, our point is to illustrate that obtaining converged, stable calculations in the EOM-IMSRG is relatively foolproof for a wide range of states, whereas the vTDA-IMSRG calculations are fraught with numerical difficulties.}. In nuclei with sub-shell closures such as $^{22}$O, the vTDA-IMSRG(2) fails to converge even for most low-lying $1$p$1$h dominant states. For this reason, we will restrict ourselves to the EOM-IMSRG(2,2) formalism in the remainder of this work.

\begin{figure}[t]
  \centering
  \includegraphics[width=0.49\textwidth]{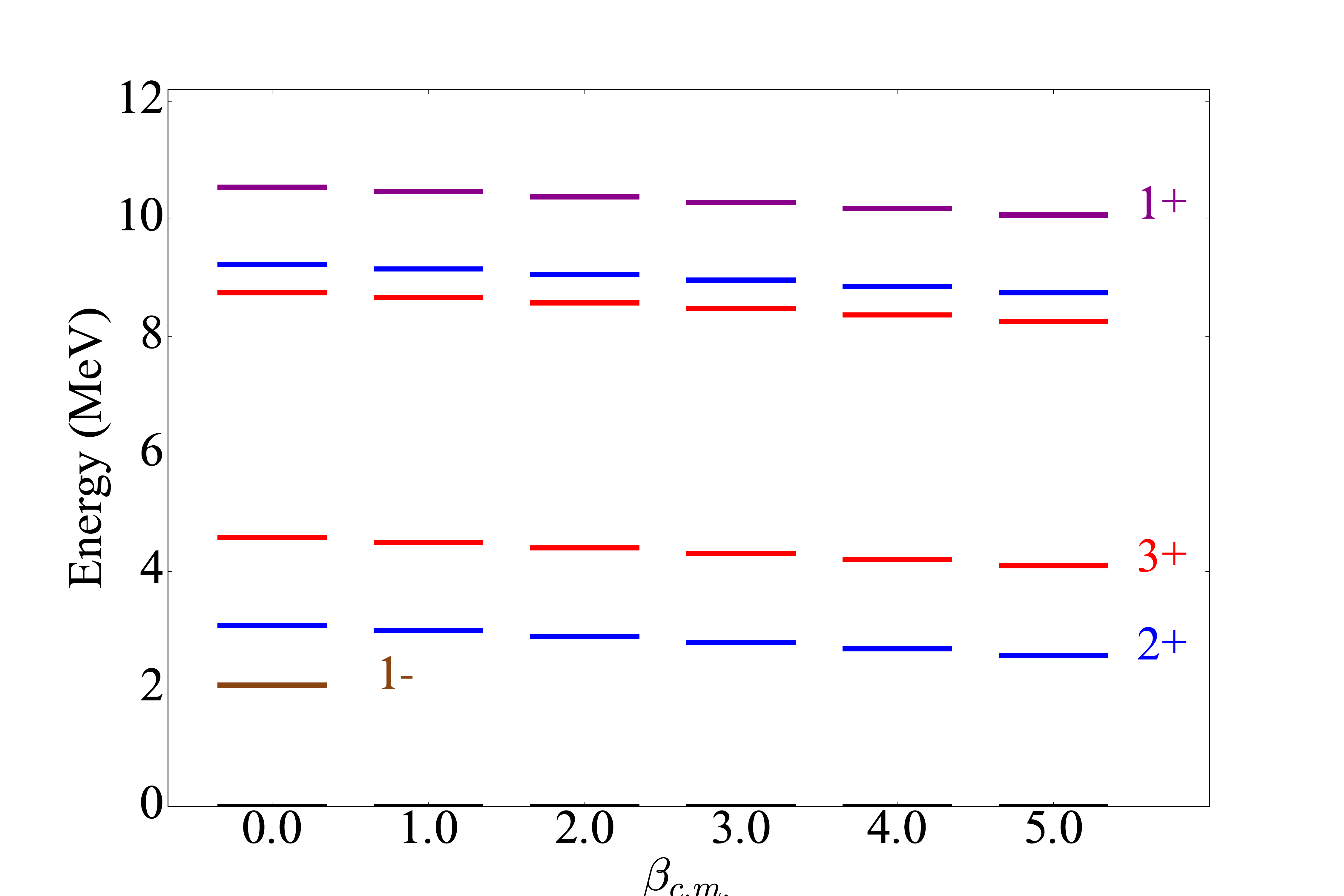}
  \caption{\label{fig:lawson_evolve}(Color online) Low lying states of $^{22}$O at $\hbar \omega = 28.0$ MeV and $e_{max}=11$ for several values of the Lawson parameter $\beta$, using the N$^3$LO (500 MeV) NN interaction of Entem and Machleidt (EM)~\cite{Entem:2003th}, softened by free-space SRG evolution to $\lambda = 3.0$ fm$^{-1}$. The COM frequency $\hbar \tilde{\omega}= 17.28$ MeV. }
\end{figure}

As discussed in Sec.~\ref{sec:systems}, spurious COM excitations are treated via the Lawson-Gloeckner method \cite{Gloeckner:1974gb}, with an augmented intrinsic Hamiltonian
\begin{equation}\label{lawson_ham} 
H  = H_{int} + \beta H_{{c}.\!{m}.\!}(\tilde{\omega})\,, 
\end{equation} 
where $\tilde{\omega}$ is determined with the method of Hagen et. al. \cite{Hagen:2009fk,Morris:2015ve}. Assuming that the intrinsic and COM wave functions factorize, the use of the Lawson term in Eq.~\ref{lawson_ham} should remove spurious excitations from the low-lying spectrum as $\beta$ is increased. An example is shown in Figure~\ref{fig:lawson_evolve} for the $1^-$ spurious state, which gets shifted out of the spectrum for non-zero values of $\beta$.   
The weak residual $\beta$ dependence of the remaining states indicates that the COM factorization is approximately satisfied for these states. We expect this factorization to improve with weaker $\beta$ dependence) as we go to higher excitation levels and larger bases, as has been empirically observed in \cite{Hagen:2009fk,Hagen:2014ve}.   
\begin{figure*}[t]
  \centering
   \includegraphics[width=0.49\textwidth]
 {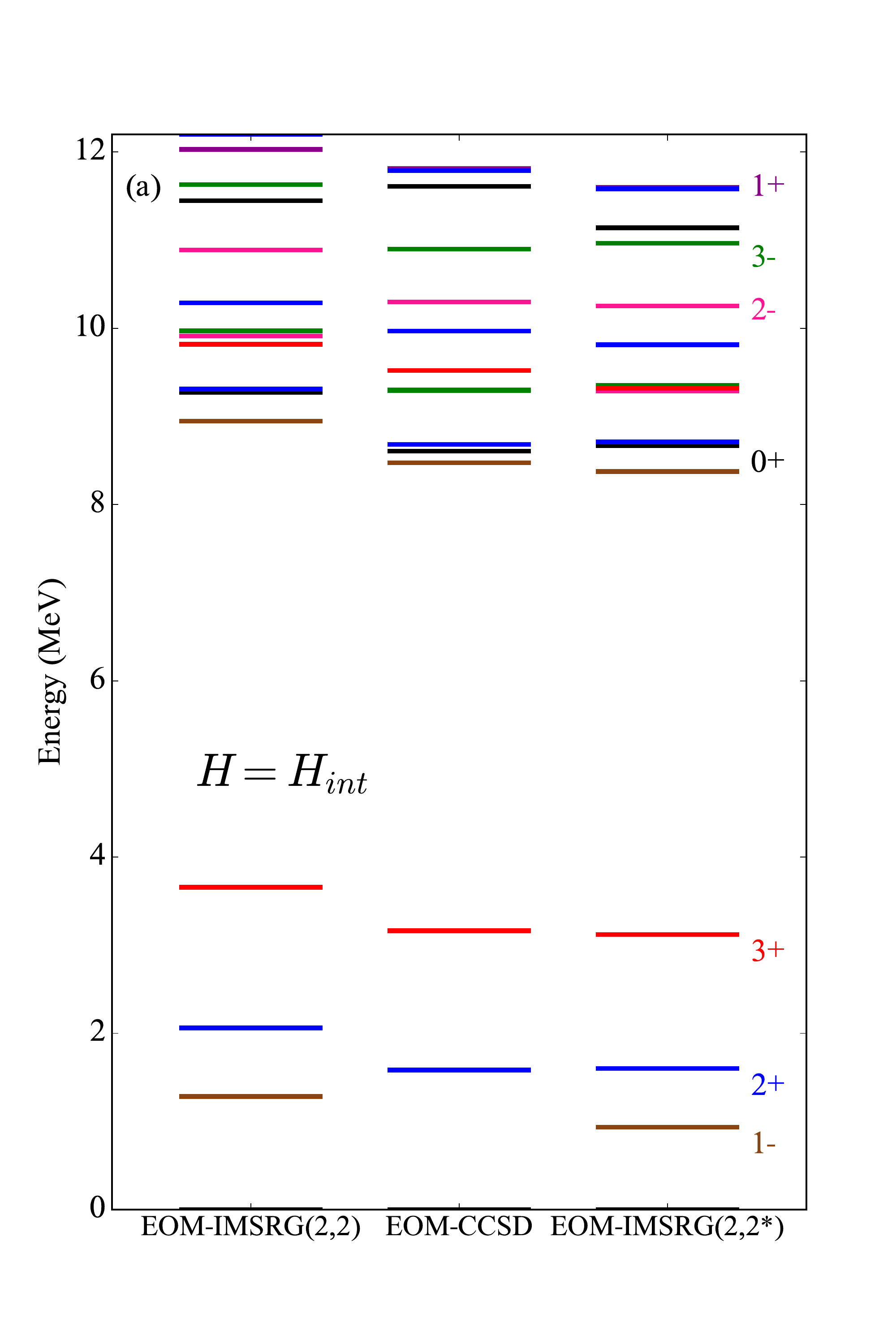}~~~%
 \includegraphics[width=0.49\textwidth]%
 {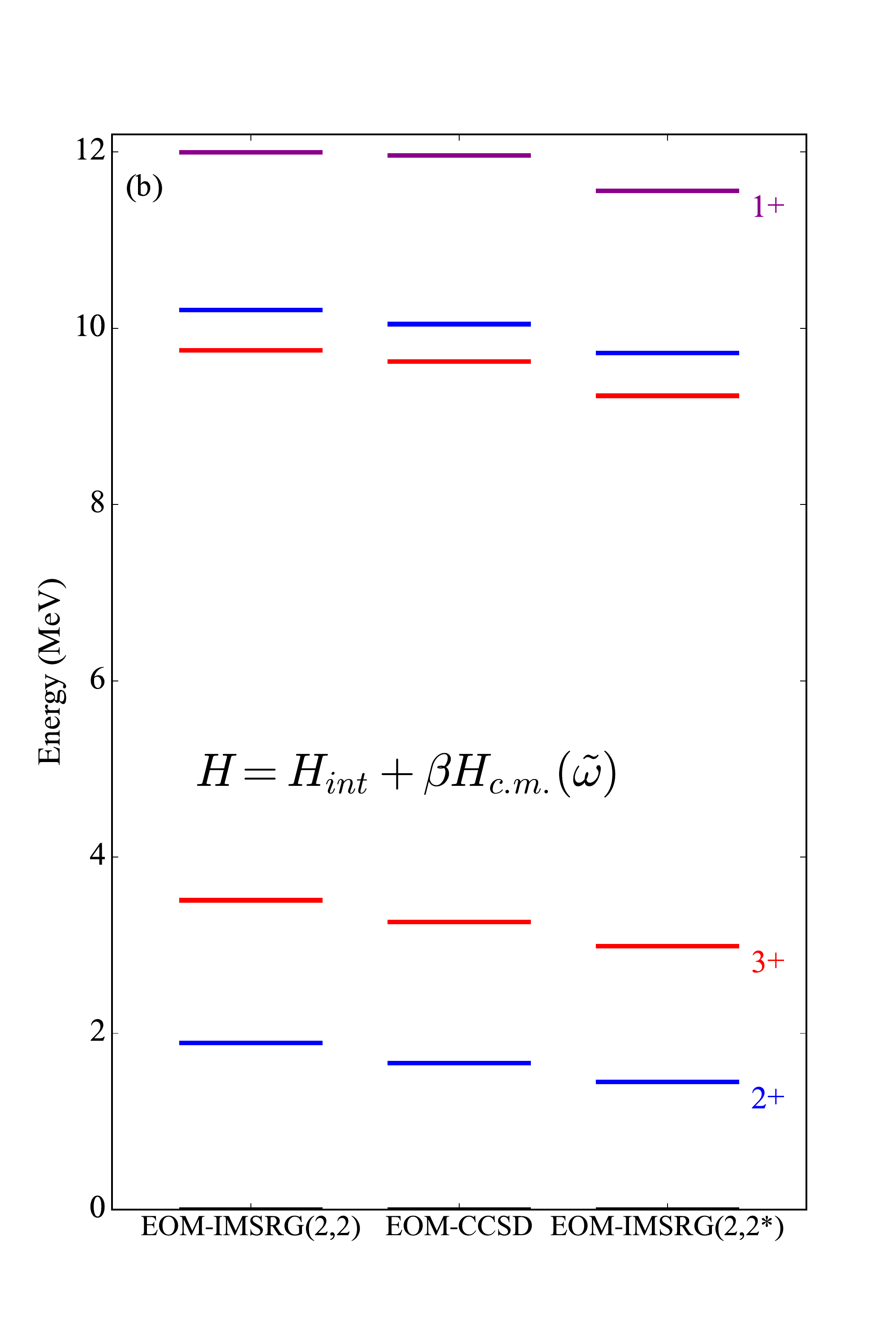}
   \caption{\label{fig:vsrg2.0_spectrum_022}(Color online) Selected excitation spectra of $^{22}$O at $\hbar \omega = 20.0$ MeV and $e_{max}=11$ using the N$^3$LO (500 MeV) NN interaction of Entem and Machleidt (EM)~\cite{Entem:2003th}, softened by free-space SRG evolution to $\lambda = 2.0$ fm$^{-1}$. The left frame shows the excitation energies calculated with the intrinsic Hamiltonian, and the right frame shows the result of adding a Lawson center-of-mass term $H=H_{int} + \beta H_{{c}.\!{m}.\!}(\tilde{\omega})$, with $\beta$ = 5.0. Different colors indicate different J$^\Pi$.  
 }
\end{figure*}

\begin{figure*}[t]
  \centering
   \includegraphics[width=0.49\textwidth]
 {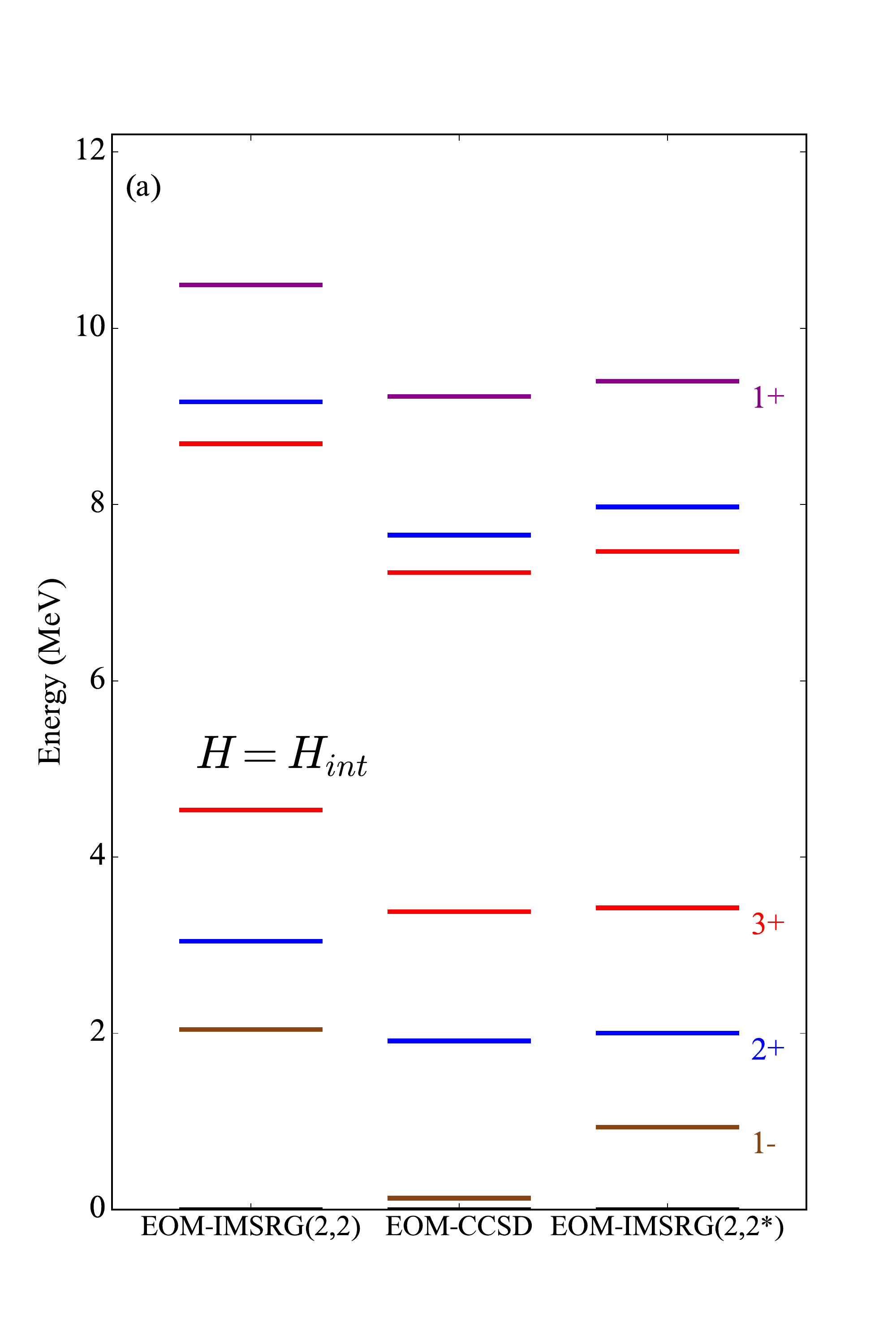}~~~%
 \includegraphics[width=0.49\textwidth]%
 {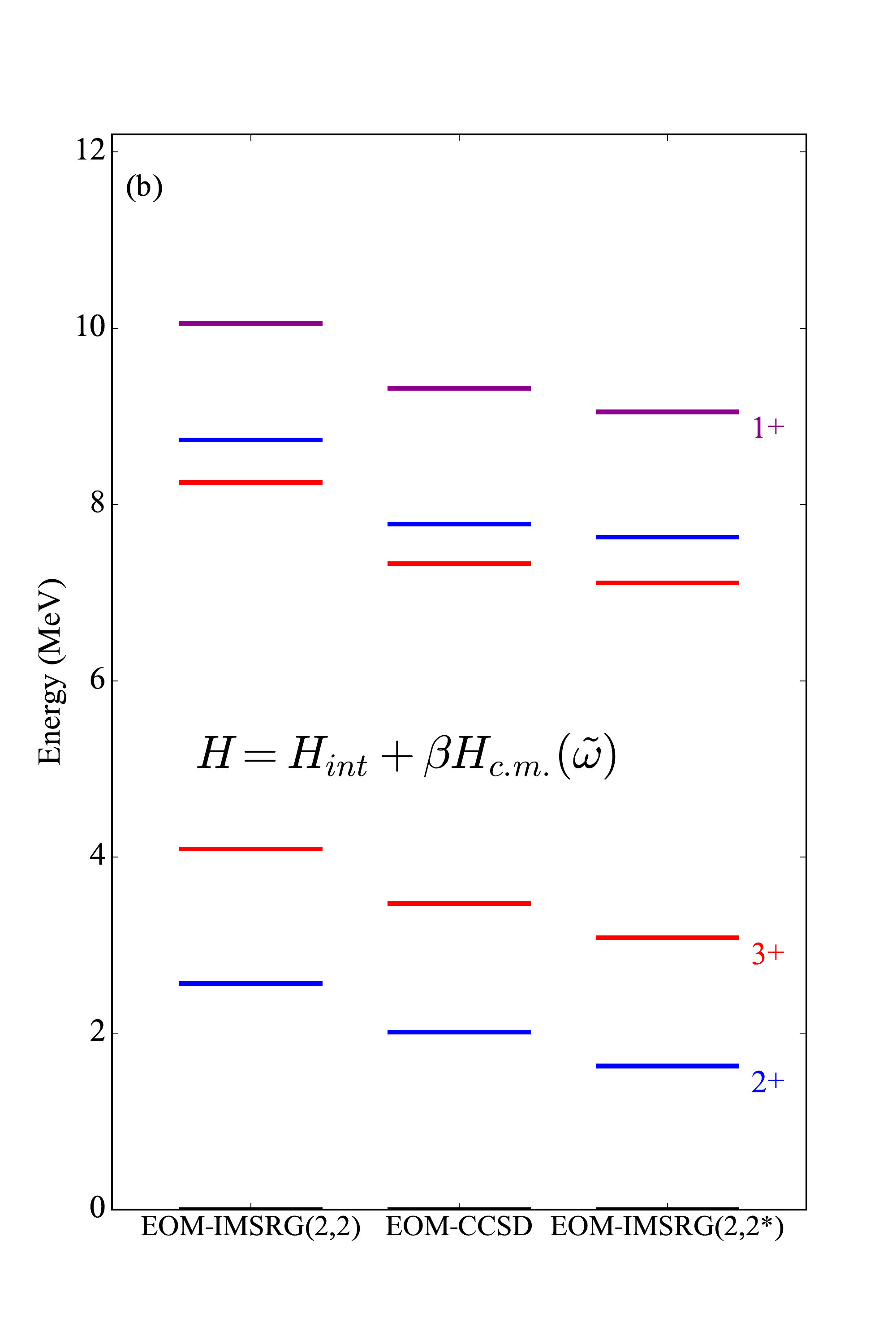}
   \caption{\label{fig:vsrg3.0_spectrum}(Color online) Selected excitation spectra of $^{22}$O at $\hbar \omega = 28.0$ MeV and $e_{max}$=11 using the N$^3$LO (500 MeV) NN interaction of Entem and Machleidt (EM)~\cite{Entem:2003th}, softened by free-space SRG evolution to $\lambda = 3.0$ fm$^{-1}$. The left frame shows the excitation energies calculated with the intrinsic Hamiltonian, and the right frame shows the result of adding a Lawson center-of-mass term $H=H_{int} + \beta H_{{c}.\!{m}.\!}$, with $\beta$ = 5.0. Different colors indicate different J$^\Pi$.  
 }
\end{figure*}

An important litmus test for the EOM-IMSRG(2,2) method will be the ability to produce results that are comparable to analogous EOM-CC calculations.  As with ground-state coupled cluster theory, EOM-CC methods originated in nuclear physics ~\cite{Emrich:1981vku,Emrich:1981ezk,Emrich:1981yfl}, but were largely ignored for many years due to convergence issues arising from the ``hard cores'' found in most NN potential models, while going on to enjoy great success in quantum chemistry~\cite{BartlettEOM,Shavitt:2009}. In recent years, EOM-CC methods have had a resurgence in nuclear physics due to the development of softer chiral EFT interactions and RG methods to soften them further, providing unprecedented access to \emph{ab initio} calculations of ground and excited state properties for medium-mass nuclei in the vicinity of closed shells~\cite{ Wloch:2005qq,Gour:2005dm,Jansen:2012ey,Hagen:2012oq,Ekstrom:2014iya,Ekstrom:2015fk}. 

Due to similar truncations being made, we expect the EOM-IMSRG(2,2) to produce results that are comparable to EOM-CC calculations truncated at single and double excitations (EOM-CCSD). In Figure~\ref{fig:vsrg2.0_spectrum_022} we show calculations of the low-lying spectra of $^{22}$O performed on the intrinsic Hamiltonian, as well as the Lawson Hamiltonian with $\beta$=5.0. 
In each panel, the left column shows the excited states calculated in the standard EOM-IMSRG(2,2) approximation, while the right column shows a slightly different approximation that we call EOM-IMSRG(2,2*). 

The latter is based on the observation that in terms of low-order MBPT content, the IMSRG(2) differs from CCSD by undercounting a class of 4th-order quadruple-excitation contributions to the correlation energy by a factor of $1/2$~\cite{Evangelista,TitusThesis}. This difference explains the empirical observation that IMSRG(2) ground-state energies tend to fall in between CCSD and CCSD(T) calculations for a wide range of single-reference systems, as the undercounting mimics the partial cancellation that occurs between   
repulsive quadruple-excitation contributions and the attractive triples corrections. In Ref.~\cite{TitusThesis}, the IMSRG(2*) approximation was developed where a class of terms which are neglected in the the strict NO2B truncation are restored, bringing the counting of the quadruple-excitation diagrams into full agreement with CCSD. In the present work, the EOM-IMSRG(2,2*) utilizes the IMSRG(2*) ground-state-decoupled Hamiltonian as input for the EOM calculation. The spectra calculated using either Hamiltonian are rather similar, with qualitative agreement between the EOM-IMSRG and EOM-CC methods for all investigated quantum numbers.

On a technical note, COM frequencies $\tilde{\omega}$ are calculated independently for the IMSRG(2) and IMSRG(2*) methods, and corresponding Lawson terms are constructed. The Lawson term is constructed in the CCSD calculations using the frequencies calculated in the IMSRG(2*) formalism, which we expect to be a good approximation given the similar perturbative content of both methods. The relevant frequencies are given in table~\ref{tab:O22_omegatilde}. The removal of spurious center-of-mass excitations is consistent in all three approaches. 

\begin{table}[b]
\centering
\caption{Values of $\hbar \tilde{\omega}$ in MeV, used in center-of-mass Hamiltonian for Lawson calculations of $^{22}$O energy spectra.}
\label{tab:O22_omegatilde}
\begin{tabular*}{0.48\textwidth}{@{\extracolsep{\fill}}ccc}
\hline \hline
  Method & $\lambda=2.0$ fm$^{-1}$  & $\lambda=3.0$ fm$^{-1}$ \\
  \hline
  IMSRG(2)  & 18.19 & 17.28 \\ 
  IMSRG(2*) & 18.05 & 17.50 \\ 
  CCSD       & 18.05 & 17.50 \\ \hline\hline
 \end{tabular*} 
\end{table}
Figure~\ref{fig:vsrg3.0_spectrum} displays a similar comparison for a ``harder'' interaction at $\lambda$ = 3.0 fm$^{-1}$. Differences between the EOM-IMSRG(2,2*) and CCSD are more notable here, but qualitative agreement is still intact. The EOM-IMSRG(2,2)  excitation energies are generally shifted up from their CCSD counterparts. This is indicative of the increasing effect of missing 4th-order quadruples for harder interactions. This is consistent with observations of increasing differences between IMSRG(2) and CCSD ground-state energies for interactions with larger $\lambda$ values~\cite{Hergert:2015awm}.

\section{Summary and Outlook} 
\label{sec:summary}
In this work we have presented two new approaches for performing \emph{ab initio} calculations of excited states in closed-shell, medium-mass nuclei. Both approaches are based on combining the IMSRG with simple many-body methods commonly used to target excited states, such as the Tamm-Dancoff approximation (TDA) and equations-of-motion (EOM) techniques. In the first method, a two-step sequential IMSRG(2) decoupling is used to drive the Hamiltonian to a form where a simple TDA calculation (i.e., diagonalization in the space of $1$p$1$h excitations) becomes exact for a subset of eigenvalues. This is accomplished by performing a second IMSRG(2) evolution after the initial ground-state decoupling has been achieved to eliminate matrix elements between the desired $1$p$1$h block and rest of the Hilbert space.  In the second approach, which we refer to as the EOM-IMSRG(2,2) method, standard equations-of-motion (EOM) techniques with single- and double-excitation operators are applied to ground-state IMSRG(2) calculations to access excited states.

We introduced two variants of the sequential decoupling approach, TDA-IMSRG(2) for the case where the entire $1$p$1$h block is decoupled, and vTDA-IMSRG(2) for the case where the particle orbital is restricted to lie in a valence shell.  The results for the sequential decoupling approaches are typically more accurate than the corresponding EOM-IMSRG(2,2) calculations when they converge, but they lack the versatility and numerical stability of the latter. This was demonstrated in detail for parabolic quantum dots in 2d, where correlations could be controlled by changing the trapping frequency $\omega$, and comparisons against exact FCI calculations could be made for sufficiently small single-particle bases.  Moreover, the EOM-IMSRG calculations are systematically improvable, as evidenced by the EOM-IMSRG(\{3\},2) quantum dot spectra that contain perturbative corrections for the neglected triple-excitation components in the EOM ladder operator. 

For calculations of $^{16,22}$O nuclei, the differences in numerical stability were even more stark, as the vTDA-IMSRG(2) approach only converged for the doubly-magic $^{16}$O nucleus, while the TDA-IMSRG(2) failed to converge at all. In contrast, all of our EOM-IMSRG calculations were free of numerical issues, with calculated spectra in good qualitative agreement with analogous EOM-CCSD calculations.   We view the qualitative agreement with the EOM-CCSD spectra as an important ``proof-of-principle'' of the EOM-IMSRG method that paves the way for more interesting applications in the near term, such as comparisons of EOM-CCSD(T) and EOM-IMSRG(\{3\},2) spectra in nuclei, calculations of consistently evolved observables such as electromagnetic strength functions and nuclear matrix elements, and extensions of the EOM-IMSRG to nuclei within 1-2 nucleons of a closed shell by generalizing the EOM ladder operator to include particle-number nonconserving terms. 

As discussed in the introduction, the Hermiticity of the EOM-IMSRG might bring some modest technical advantages over analogous EOM-CC methods. However, the ultimate advantage of the EOM-IMSRG lies in an eventual multi-reference formulation, which, owing to the relative simplicity (both technical and formal) of the MR-IMSRG compared to MR-CC, has the potential to extend calculations well into the open-shell regime, while avoiding the ``curse of dimensionality'' that ultimately limits large-scale shell model calculations.


\section{Acknowledgments}
We thank Gaute Hagen, Heiko Hergert, Morten Hjorth-Jensen, Gustav Jansen, and Ragnar Stroberg for useful discussions, and we thank Gaute Hagen for the use of his EOM-CCSD code. This work was supported in part by the National Science Foundation under Grant No. PHY-1404159, and by the NUCLEI SciDac Collaboration under DOE Grant No. DE-SC000851. This work was supported by the Office of Nuclear Physics, U.S. Department of Energy, under Grant No. DE-SC0008499 (NUCLEI SciDAC Collaboration) Oak Ridge National Laboratory is supported by the Office of Science of the Department of Energy under Contract No. DE-AC05-00OR22725.

%

\end{document}